\documentclass[a4paper,11pt]{article}
\pdfoutput=1
\usepackage{jcappub}
\usepackage[T1]{fontenc}
\usepackage{amsmath,amssymb}
\usepackage{color}
\usepackage{subfigure}
\usepackage{graphicx}
\usepackage[dvipsnames]{xcolor}

\title{\boldmath Cosmological dynamics of mimetic gravity}

\author[a,b]{Jibitesh Dutta}
\author[c,d]{Wompherdeiki Khyllep}
\author[e,f,g]{Emmanuel N. Saridakis}
\author[h,i]{Nicola Tamanini}
\author[j,k]{and Sunny Vagnozzi}

\affiliation[a]{Mathematics Division, Department of Basic Sciences and Social Sciences,
North
Eastern Hill University,
NEHU Campus, Shillong, Meghalaya 793022, India}
\affiliation[b]{Inter University Centre for Astronomy and Astrophysics(IUCAA), Pune 411 
007,
India}
\affiliation[c]{Department of Mathematics, North Eastern Hill University, NEHU Campus,
Shillong,
Meghalaya 793022, India}
\affiliation[d]{Department of Mathematics, St. Anthony's College, Shillong, Meghalaya
793001, India}
\affiliation[e]{Chongqing University of Posts \& Telecommunications, Chongqing,
400065,China}
\affiliation[f]{Department of Physics, National Technical University of Athens, Zografou
Campus GR
157 73, Athens, Greece}
\affiliation[g]{CASPER, Physics Department, Baylor University, Waco, TX 76798-7310, USA}
\affiliation[h]{Max-Planck-Institut f\"{u}r Gravitationsphysik, Albert-Einstein-Institut,
Am M\"{u}hlenberg 1, 14476 Potsdam-Golm, Germany}
\affiliation[i]{Laboratoire Astroparticule et Cosmologie, CNRS UMR 7164, Universit\'e
Paris-Diderot,
 10 rue Alice Domon et L\'eonie Duquet, 75013 Paris, France}
\affiliation[j]{The Oskar Klein Centre for Cosmoparticle Physics, Department of Physics,
Stockholm
University, SE-106 91 Stockholm, Sweden}
\affiliation[k]{The Nordic Institute for Theoretical Physics (NORDITA),
Roslagstullsbacken
23, SE-
106 91 Stockholm, Sweden}

\emailAdd{jibitesh@nehu.ac.in}
\emailAdd{sjwomkhyllep@gmail.com}
\emailAdd{Emmanuel\_Saridakis@baylor.edu}
\emailAdd{nicola.tamanini@aei.mpg.de}
\emailAdd{sunny.vagnozzi@fysik.su.se}

\abstract{
We present a detailed investigation of the dynamical behavior of mimetic gravity with
a general potential for the mimetic scalar field. Performing a phase-space and stability
analysis,
we show that the scenario at hand can successfully describe the thermal history of
the universe, namely the successive sequence of radiation, matter, and dark-energy eras.
Additionally, at late times the universe can either approach a de Sitter solution, or a
scaling accelerated attractor where the dark-matter and dark-energy density parameters
are of the same order, thus offering an alleviation of the cosmic coincidence problem.
Applying our general analysis to various specific potential choices, including the
power-law and the exponential ones, we show that mimetic gravity can be brought into good
agreement
with
the observed behavior of the universe. Moreover, with an inverse square potential we find
that mimetic gravity offers an appealing unified cosmological scenario where both dark
energy and
dark matter are characterized by a single scalar field, and where the cosmic coincidence
problem is alleviated.
}


\begin{document}
\maketitle
\flushbottom

\section{Introduction}
\label{sec:introduction}

The dark universe picture, astonishingly confirmed by a variety of experiments and
surveys
in the past years
\citep{riess:1998cb,perlmutter:1998np,abbott:2017wau},
paints a rather peculiar picture of the universe we live in. In particular, observations
indicate that around 25\% of the energy  budget of the universe corresponds to the dark
matter sector, while around 70\% constitutes the dark energy one \citep{ade:2015xua},
with both sectors being relatively unknown for the moment. Concerning dark matter, there
have been numerous attempts to attribute it to specific candidates \cite{Bertone:2004pz},
such as a
weakly interacting massive particle~\cite{Roszkowski:2017nbc}, while several theories
instead posit
the existence of additional particles and forces beyond those of the Standard Model
(e.g.~\cite{ArkaniHamed:2008qn,CyrRacine:2012fz,Petraki:2014uza,
Foot:2014uba,Foot:2014osa,Foot:2016wvj}).

On the other hand,
the description of dark energy is more uncertain, since if it does not correspond to the
simplest
choice
of a cosmological constant (which nonetheless is problematic from a theoretical point of
view~\cite{Weinberg:1988cp,Sahni:2002kh}), then there are essentially two
ways to explain it.
Firstly, assuming that
general relativity correctly describes gravity on both galactic and cosmological scales,
one can attribute dark energy to various fields or exotic matter
and energy components (for reviews see \cite{Copeland:2006wr,Cai:2009zp}).  However, a
second intriguing possibility is that general relativity might not in fact be the correct
theory of gravity on all scales, thus paving the road for the wide plethora of modified
gravity theories (for reviews see
\citep{nojiri:2006ri,nojiri:2010wj,tsujikawa:2010zza,capozziello:2011et,cai:2015emx,
nojiri:2017ncd}). Modified theories of gravity are theoretically and observationally
appealing, and
ever-increasingly sensitive and precise upcoming surveys provide the exciting
possibility of robustly testing these theories against observational data.

One interesting class of modified gravity is the mimetic gravitational construction,
which has been proposed in 2013 \citep{chamseddine:2013kea} and has received considerable
attention since then. In its original formulation the mimetic theory of gravity can be
obtained starting from general relativity, by isolating the conformal degree of freedom
of gravity in a covariant fashion through a re-parametrization of the physical metric  in
terms of an auxiliary metric and a mimetic field. It is then shown that the resulting
gravitational field equations feature an additional term sourced by the mimetic field,
which can be interpreted as the contribution of a pressureless perfect fluid. It is
furthermore shown that, on a Friedmann-Robertson-Walker (FRW) background, this fluid
behaves precisely as a dust component. Thus, the mimetic field can mimic cold dark matter
on cosmological scales, a feature that gave the theory its name.

Several extensions of the basic mimetic gravity have been proposed and studied in
detail in the literature (for a review see  \citep{sebastiani:2016ras}). The earliest
extension proposed, motivated by possible caustic
instabilities, was based on a Proca-like vector field \citep{barvinsky:2013mea}.
Motivated instead by possible ghost instabilities, a mimetic tensor-vector-scalar theory
was presented in~\citep{chaichian:2014qba}.
Another interesting generalization envisions the addition of a potential $V(\phi)$ for
the mimetic scalar field \citep{chamseddine:2014vna}. It was shown that in such a way it
is possible to mimic any given cosmological background evolution by a suitable choice of
the potential. Other   extensions have  been carried out adding higher-order curvature
invariants, motivated by the fact that such terms usually
appear in the low-energy effective gravitational action when quantum or stringy
corrections are taken into account. The example \textit{par excellence} in this sense is
mimetic $F(R)$ gravity \citep{nojiri:2014zqa,leon:2014yua}, where the same procedure
leading to mimetic gravity from general relativity is applied to the $F(R)$ gravitational
framework. In the same spirit, other modifications to the curvature sector of mimetic
action have led to different proposed extensions (see e.g.~\cite{momeni:2014qta} for one 
of the 
first extensions), such as mimetic
$f(R,T)$~\citep{momeni:2015gka}, mimetic $f(G)$~\citep{astashenok:2015haa}, mimetic
$f(R,\phi)$~\citep{myrzakulov:2015qaa}, mimetic covariant Ho\v{r}ava-like
gravity~\citep{myrzakulov:2015nqa,cognola:2016gjy}, mimetic
Horndeski gravity~\citep{arroja:2015wpa,cognola:2016gjy}, mimetic Galileon
gravity~\citep{haghani:2014ita,rabochaya:2015haa}, non-local mimetic $F(R)$
gravity~\citep{myrzakulov:2016hrx},
unimodular-mimetic  $F(R)$
gravity~\citep{odintsov:2016imq}, and mimetic Born-Infeld
gravity~\citep{bouhmadi-lopez:2017lbx,chen:2017ify}. The impact of higher-derivative
terms in the mimetic gravity
action was then studied in \citep{capela:2014xta,mirzagholi:2014ifa}, as well as
in~\citep{ramazanov:2015pha,koshelev:2015ush,paston:2017das}, and motivated by
instability issues in the recent works
\citep{hirano:2017zox,zheng:2017qfs,cai:2017dxl,takahashi:2017pje,
gorji:2017cai}. Other extensions of mimetic gravity   include bi-scalar mimetic
models~\citep{saridakis:2016mjd}, vector-tensor mimetic
gravity~\citep{kimura:2016rzw}, braneworld mimetic gravity~\citep{sadeghnezhad:2017hmr},
and extensions implementing the limiting curvature hypothesis and hence constructed to
resolve cosmological singularity issues~\citep{chamseddine:2016uef,chamseddine:2016ktu}.
Finally, models in which the mimetic field is non-minimally coupled to
matter~\citep{vagnozzi:2017ilo} and baryon
number~\citep{shen:2017rya} currents (in the latter case in order to enable baryogenesis)
have been also studied.

Apart from the basic construction of a mimetic model one may analyze the
perturbations and the stability of the
theory~\citep{chamseddine:2014vna,capela:2014xta,mirzagholi:2014ifa,matsumoto:2015wja,
momeni:2015gka,
ramazanov:2015pha,koshelev:2015ush,paston:2017das,
arroja:2015yvd,
cognola:2016gjy,ramazanov:2016xhp, odintsov:2016oyz,firouzjahi:2017txv,
yoshida:2017swb,hirano:2017zox,zheng:2017qfs,cai:2017dxl,takahashi:2017pje,
gorji:2017cai}. Several recent works have pointed out that mimetic gravity
might suffer from ghost instabilities (although not of the type associated to higher
derivative instability, such as the Ostrogradsky ghost), as well as gradient
instabilities, in the scalar and/or
tensor sectors \citep{barvinsky:2013mea,chaichian:2014qba,malaeb:2014vua,
haghani:2015zga,langlois:2015skt,ramazanov:2016xhp,achour:2016rkg,
ijjas:2016pad,kluson:2017iem, firouzjahi:2017txv,yoshida:2017swb,
arroja:2015yvd,odintsov:2016oyz,hirano:2017zox,zheng:2017qfs, cai:2017dxl,
takahashi:2017pje,gorji:2017cai}. Although currently a lack of consensus
regarding
these issues persists, it is worth pointing out that if mimetic gravity does indeed
suffer from instabilities then there are ways to rescue the theory, e.g.~by directly
coupling higher derivatives of the mimetic field to curvature.

Mimetic gravity can have interesting cosmological applications. Numerous works have
investigated in detail the cosmological
phenomenology of the above constructions, such as inflation, late-time acceleration, and
unified universe evolution including the  intermediate radiation
and matter eras
\citep{chamseddine:2014vna,nojiri:2014zqa,saadi:2014jfa,leon:2014yua,
haghani:2015iva,
matsumoto:2015wja,momeni:2015gka,astashenok:2015haa,myrzakulov:2015qaa,
arroja:2015wpa,haghani:2014ita,rabochaya:2015haa,hammer:2015pcx,
myrzakulov:2015nqa,cognola:2016gjy,
myrzakulov:2016hrx,odintsov:2016imq,nojiri:2016vhu,
hirano:2017zox,zheng:2017qfs,cai:2017dxl,takahashi:2017pje,
gorji:2017cai,baffou:2017pao,paston:2017das,
bouhmadi-lopez:2017lbx,
saridakis:2016mjd,kimura:2016rzw,sadeghnezhad:2017hmr,chamseddine:2016uef,
chamseddine:2016ktu,vagnozzi:2017ilo,
shen:2017rya}. In particular~\cite{raza:2015kha} focused on cosmological attractors.
Additionally, in the above framework  one can study black hole solutions
\citep{deruelle:2014zza,myrzakulov:2015sea,addazi:2017cim}, compact ``stellar'' objects
which solve the
Tolman-Oppenheimer-Volkoff equations  \citep{momeni:2015fea}, quark and
neutron stars~\citep{astashenok:2015qzw}, wormholes~\citep{myrzakulov:2015kda},
modifications to the local gravitational potential which can potentially
explain the flatness of rotation curves
\citep{myrzakulov:2015kda,vagnozzi:2017ilo},
and gravitational focusing of mimetic matter~\citep{babichev:2016jzg}.
Finally, confronting the theory with observations, one can use data from
large-scale structure~\citep{arroja:2017msd}  and gravitational waves
\citep{ezquiaga:2017ekz,sakstein:2017xjx,baker:2017hug,langlois:2017dyl,visinelli:2017bny,
mimeticgws1,mimeticgws2} surveys.

It has also been realized that mimetic gravity is intimately connected to a number of
other well-known theories of modified gravity. Perhaps the most remarkable connection is
the one which considers mimetic gravity to appear in the infrared limit
of the projectable version of Ho\v{r}ava-Lifshitz gravity. This correspondence has been
formally proven in~\citep{ramazanov:2016xhp} and shows that mimetic gravity can be viewed
as the low-energy limit of a (Lorentz-violating) theory of quantum gravity.
Intimate relations between mimetic gravity and other  theories   include connections to
the scalar Einstein-Aether theory
\citep{haghani:2014ita,jacobson:2015mra},
Ho\v{r}ava-like theories with dynamical diffeomorphism invariance
breaking~\citep{myrzakulov:2015nqa},
degenerate higher-order scalar-tensor theories
beyond Horndeski~\citep{achour:2016rkg}, singular Brans-Dicke theory
\citep{hammer:2015pcx}, non-commutative geometry~\citep{chamseddine:2014nxa}, effective
implementations of the limiting curvature
criterion~\citep{chamseddine:2016uef,chamseddine:2016ktu,
yoshida:2017swb,chinaglia:2017wim,gorji:2017cai}, as well as to other more exotic
theories
of gravity~\citep{guendelman:2015jii,benisty:2017eqh,kopp:2016mhm,
ali:2015ftw,
paston:2017das,benisty:2017lmt}.

In the present work we are interested in performing a complete dynamical analysis of
the cosmological evolution in mimetic gravity, since although mimetic cosmology has been
extensively studied, the dynamical behavior of these solutions has not been
explored in detail (apart from the sub-class of mimetic $F(R)$ gravity 
\cite{leon:2014yua}). 
Dynamical systems analysis provides a
very powerful tool in the study of the asymptotic behavior as well as of the complete
cosmological dynamics of a given cosmological model
\citep{coley:2003mj,Leon2011,boehmer:2014vea,DSreview}.
This phase-space and stability examination allows to bypass the non-linearities of
the cosmological equations, and  obtain a description of the global dynamics
independently of the initial conditions of the universe, connecting critical points to
epochs of the evolutionary history which are of particular relevance. In particular, a
late-time period
of accelerated expansion would typically correspond to a late-time attractor, whereas
epochs of radiation and  matter domination typically correspond to saddle points.
With the full machinery of dynamical systems is indeed possible to investigate the
complete dynamics of any cosmological model, provided suitable dynamical variables can be
identified. Hence, these powerful methods have been extensively applied to
analyse the evolution of several cosmological models, including many modified gravity
scenarios (see
e.g.~\citep{Copeland:1997et,Leon:2009rc,Xu:2012jf,Leon:2012mt,Leon:2013qh,Kofinas:2014aka,
Carloni:2015bua, Carloni:2015lsa, Boehmer:2015kta, Boehmer:2015sha, Dutta:2016bbs,
Tamanini:2016klr,Dutta:2017kch,Dutta:2017wfd,Zonunmawia:2017ofc,Carloni:2017ucm,
Carloni:2004kp,Carloni:2007eu,Hohmann:2017jao} for
some
recent works).

This paper is organized as follows: In section \ref{sec:mimetic_gravity} we provide a
review of mimetic gravity, and we apply it within a cosmological framework. In section
\ref{sec:dynamical_system_analysis} we perform a detailed phase-space and stability
analysis of mimetic cosmology for a general potential, and then we specify the
investigation in the case of various specific potentials, amongst others for the
power-law and the exponential ones. In section \ref{sec:cosmological_implications} we
discuss the cosmological implications of the obtained results. Finally, section
\ref{sec:conclusion} is devoted to the conclusions.

\section{Mimetic gravity and cosmology}
\label{sec:mimetic_gravity}

  In this section we briefly review the current status of mimetic gravity. For further
details, we refer the reader to the recent review \citep{sebastiani:2016ras}.
In its original formulation the mimetic theory of gravity can be obtained starting from
general relativity. In particular, isolating the conformal degree of freedom of
gravity in a covariant fashion  by parametrizing the physical metric $g_{\mu \nu}$ in
terms of an auxiliary metric $\tilde{g}_{\mu \nu}$ and the mimetic field $\phi$, one can
write \citep{chamseddine:2013kea}
\begin{eqnarray}
g_{\mu \nu} = \tilde{g}_{\mu \nu}\tilde{g}^{\alpha
\beta}\partial_{\alpha}\phi\partial_{\beta}\phi \, .
\label{mimetic}
\end{eqnarray}
This mimetic parametrization makes it clear that the physical metric is invariant under
conformal transformations of the auxiliary metric.
It is easy to show that, for consistency, the following condition on the gradient of the
mimetic field has to be satisfied \citep{chamseddine:2013kea}
\begin{eqnarray}
g^{\mu \nu}\partial_{\mu}\phi\partial_{\nu}\phi = 1 \, .
\label{partialphi}
\end{eqnarray}

The consistency condition (\ref{partialphi}) can be
implemented at the level of the action through a Lagrange multiplier constraint as
\citep{chamseddine:2014vna}
\begin{eqnarray}
I = \int d^4x \sqrt{-g} \left [ \frac{R}{2 \kappa^2} + \lambda \left(
\partial^\mu\phi\partial_\mu\phi -1 \right) + \mathcal{L}_m \right ] \, ,
\label{actionlagrangemultiplier}
\end{eqnarray}
where $\kappa^2$ is the gravitational constant, $R$ is the Ricci scalar, and $\mathcal{L}_
m$ is the usual standard-model matter  Lagrangian,
and thus variation of the action with respect to the Lagrange multiplier field $\lambda$
enforces the validity of the constraint (\ref{partialphi}) (actions
featuring Lagrange-multiplier constrained scalar fields are in general   being used in
the literature, see for instance
\citep{lim:2010yk,gao:2010gj,capozziello:2010uv}).

As we mentioned in the Introduction,  a simple extension of the original mimetic
construction (\ref{actionlagrangemultiplier}) is to include a potential for the mimetic
field. In particular, one writes  \citep{chamseddine:2014vna}
\begin{equation}
   I = \int d^4x \sqrt{-g} \left[ \frac{R}{2 \kappa^2} -\frac{\lambda}{2} \left(
\partial^\mu\phi\partial_\mu\phi -1 \right) + V(\phi) + \mathcal{L}_m \right] \,,
    \label{eq:action}
\end{equation}
where  $V(\phi)$ is the self-interacting scalar field potential, and where the factor of
$1/2$ in front of the Lagrange multiplier $\lambda$ is introduced for convenience.
Such a model has been shown to provide an economical way of reproducing a number of
simple and well-motivated cosmological scenarios, relevant for both early- and late-time
cosmology, without the need for neither an explicit dark matter nor a dark energy
fluid~\citep{chamseddine:2014vna}. Moreover, one can obtain   late-time accelerating
solutions, early-time inflationary states, bouncing solutions, etc. It is interesting to
mention that the above model is consistent with the latest observation of
the gravitational wave event GW170817 from a binary neutron star inspiral with
an electromagnetic counterpart event~\citep{TheLIGOScientific:2017qsa}, as the
corresponding  propagation speed of tensor perturbations is identically equal to the
speed of light~\citep{ezquiaga:2017ekz,mimeticgws1,mimeticgws2}.

The equations of motion of the theory can be obtained by varying the action with
respect to the physical metric, however taking into account its dependence on the
auxiliary metric and the mimetic field. Hence, variation of the action \eqref{eq:action}
with respect to the metric gives
\begin{equation}
\frac{1}{\kappa^2} G_{\mu\nu}=\lambda \partial_\mu \phi\partial_\nu \phi+g_{\mu\nu}
V(\phi)+T_{\mu \nu}, \label{eq:field_eq}
\end{equation}
where $G_{\mu\nu}$ is the Einstein tensor and $T_{\mu \nu}$ is the standard-model matter
energy-momentum
tensor. On the other hand, as we mentioned, variation with respect to the Lagrange
multiplier indeed yields condition  \eqref{partialphi}.
 Taking the trace of equation (\ref{eq:field_eq}) we find the Lagrange multiplier to be
\begin{equation}
\lambda=\left(\frac{G}{\kappa^2}-T-4V\right),
\label{traceeq}
\end{equation}
%
where $G$ and $T$ are the traces of the Einstein tensor and the matter  energy-momentum
tensor respectively. Finally, variation of the action \eqref{eq:action} with respect to
the mimetic field $\phi$ gives
\begin{equation}
\nabla^{\mu}\left(\lambda\partial_{\mu}\phi \right)+\frac{dV}{d\phi}=0, \label{eq:KG_eq}
\end{equation}
an equation that can alternatively be derived taking the
covariant derivative  of   Eq.~\eqref{eq:field_eq} and employing the Bianchi
identity $\nabla^\mu G_{\mu\nu}=0$ together with the conservation equation $\nabla^\mu
T_{\mu\nu}=0$.

It is   worth mentioning that one can substitute the value of $\lambda$ obtained from 
Eq.~\eqref{traceeq} into the other field equations. This operation would eliminate 
$\lambda$ from the dynamics and the resulting system of equations would involve in its 
place some combination of the degrees of freedom of the metric, the matter fluid and the 
mimetic field. In practice one does not gain any advantage by this operation 
since no additional dynamics is obtained, and computationally one ends up dealing with 
more complex expressions. Moreover, this operation would lead to the field equations of 
the original formulation of mimetic gravity theory without the Lagrange's multiplier 
$\lambda$ and this would invalidate the very purpose of taking action 
\eqref{actionlagrangemultiplier} (for details see \cite{sebastiani:2016ras}). For these 
reasons in what follows we prefer to explicitly work with $\lambda$ rather than 
eliminating it for more complicated expressions.

From equations (\ref{eq:field_eq})-(\ref{eq:KG_eq}) it is clear that the
gravitational field equations in mimetic gravity differ from those of general relativity
by the presence of an extra source term which mimics a perfect fluid. In the original
version of the theory,  where the potential is absent, this extra fluid has energy
density $\rho_f=\frac{G}{\kappa^2}-T$, four-velocity $\partial_{\mu}\phi$ and pressure
$p_f=0$. The
fact that $p_f=0$ suggests that this extra term can play the role
of a pressureless fluid and hence mimic a dust-matter component on cosmological scales.
Therefore, the construction at hand can mimic cold dark matter on cosmological scales,
bypassing the need for an additional dark matter component, that is why it is
called ``mimetic gravity''.


The foremost questions to be addressed is why did the seemingly innocuous
reparametrization of the physical metric in Eq.~(\ref{mimetic}) lead to different
equations of motion compared to general relativity. Early attempts to address this 
question 
identified the dark
matter degree of freedom as arising from gauging  the local Weyl invariance of the
theory \citep{barvinsky:2013mea}, whereas other early works explained the different
equations of motion in terms of variation of the action over a restricted class of
functions which results in a broader freedom in the dynamics of the
theory~\citep{golovnev:2013jxa}. Nowadays it is well understood that the reason
behind the fact that the equations of motion of mimetic gravity
differ from those of general relativity, is to be sought in the fact that the former is
related to the latter via a singular (i.e.~non-invertible) disformal transformation.
Recalling that general relativity satisfies diffeomorphism invariance allows
one to reparametrize the physical metric in terms of an auxiliary metric and a scalar
field through what is known as a disformal transformation \citep{bekenstein:1992pj}. An
invertible disformal transformation returns a theory which is equivalent to general
relativity. On the other hand, a non-invertible transformation modifies the 
dynamics of 
the theory. The
reparametrization
of Eq.~(\ref{mimetic}) falls within this category, thus explaining why the equations of
motion  are different from those of general relativity
\citep{deruelle:2014zza,yuan:2015tta,domenech:2015hka,deffayet:2015qwa,
arroja:2015wpa,domenech:2015tca,
achour:2016rkg,saitou:2016lvb,celoria:2016vul,ezquiaga:2017ner,takahashi:2017zgr,
crisostomi:2017ugk}.

Let us now apply mimetic gravity in a cosmological framework. We consider a flat FRW
universe (with the $(+,-,-,-)$ convention)
\begin{equation}
    ds^2 = dt^2 - a(t)^2 ( dx^2 + dy^2 + dz^2 ) \,.
\end{equation}
Assuming the scalar field to be spatially homogeneous, i.e.~$\phi(t)$,
the constraint
(\ref{partialphi}) yields
\begin{equation}
    \dot\phi^2 = 1 \,,
\end{equation}
which, choosing $\dot\phi>0$, upon integration implies
\begin{equation}
    \phi = t \,,
    \label{eq:phi_eq_t}
\end{equation}
where the integration constant has been set to zero for convenience.
The cosmological equations \eqref{eq:field_eq} and \eqref{eq:KG_eq} will then read
\begin{gather}
    3H^2 = \kappa^2 \left( \rho + \lambda + V \right) \,, \label{eq:cosmo_1} \\
    3H^2 + 2\dot H = - \kappa^2 \left(p - V \right) \,, \label{eq:cosmo_2}  \\
    \dot\lambda + 3H \lambda + \frac{d V}{d\phi} = 0 \,, \label{eq:cosmo_3}
\end{gather}
with $H=\dot{a}/a$ the Hubble function and with over-dots denoting differentiation with
respect to $t$. In the above equations we have considered as usual that the
standard-model
matter
energy-momentum tensor corresponds to a
perfect fluid of energy density $\rho$ and pressure $p$. In what follows we will assume a
linear equation of state (EoS) for the matter fluid, namely $p =w \rho$. %
Thus, the equations close by the consideration of the matter conservation equation
\begin{equation}
    \dot\rho + 3H \left( \rho + p \right) = 0 \,. \label{eq:rho_consrv}
\end{equation}
Note that Eqs.~\eqref{eq:cosmo_1}-\eqref{eq:rho_consrv} are not independent, since one 
can obtain any one of  \eqref{eq:cosmo_2}-\eqref{eq:rho_consrv} from the remaining two 
along with the constraint \eqref{eq:cosmo_1} and the condition \eqref{eq:phi_eq_t}.
This implies that in order to investigate the dynamics of the system, only an independent 
subset of these equations needs to be considered. In fact, to derive the dynamical system 
equations in Sec.~\ref{sec:dynamical_system_analysis}, we 
will start from an independent subset of them following the standard procedure commonly 
used in dynamical systems applications in cosmology \cite{DSreview}.
Nevertheless, for the sake of completeness in this section we present all   
cosmological equations that  directly follow from the variation of the mimetic 
gravity action \eqref{actionlagrangemultiplier}.

Note moreover that due to \eqref{eq:phi_eq_t} the potential $V(\phi)$ can be
considered as a function of $t$, namely $V(t)$.

We close  this section by mentioning that in the case where the potential is absent,
equations (\ref{eq:cosmo_1}) and (\ref{eq:cosmo_2}) are nothing but the usual
cosmological equations with a new non-relativistic matter
component given by the Lagrange multiplier $\lambda$, which can indeed be used to model
dark matter. In particular, from  (\ref{eq:cosmo_3}) we can clearly see that
within an FRW background, the energy density of the extra fluid decays with the
scale factor  as $a^{-3}$, precisely as expected for a dust component. Therefore, the
construction at hand can mimic cold dark matter on cosmological scales, bypassing the
need for an additional dark matter component, that is why it is
called ``mimetic gravity''.

On the other hand,  the addition of the potential term
$V(\phi)$ is considered in order to have a mechanism to additionally describe the
late-time
accelerated expansion (since without a potential this cannot be achieved).
When the potential $V(\phi)$ does not vanish Eq.~\eqref{eq:cosmo_3} 
implies that the energy density corresponding to $\lambda$ is not conserved, due to the 
interaction with the scalar field $\phi$. This implies in particular that instead of 
following the standard $a^{-3}$ scaling, $\lambda$ will evolve with a more general 
dynamics whenever the derivative of the scalar field potential is not negligible.
Note that $\lambda$ can nevertheless still be identified with the dark matter component 
at cosmological scales.
In fact, the dark matter energy density scales as $a^{-3}$ only in the absence of an 
interaction with dark energy, but whenever an interaction is present this is no longer 
true. This happens in general in every model of interacting  dark energy (see for 
example \cite{bolotin:2013jpa} or the models considered in 
\cite{amendola:1999er,wang:2016lxa,Nunes:2016dlj}).
In our case, when the potential is zero there is no interaction between dark matter and  
dark energy (since there is no  dark energy at all) and dark matter scales as 
$a^{-3}$; however if the potential is non-zero then an implicit interaction between dark 
matter (namely $\lambda$) and  dark energy (namely $\phi$) is present and the $a^{-3}$  
evolution is modified. Note however that when dark matter dominates then the  dark-energy 
energy density (and thus $V(\phi)$) is effectively zero and thus the interaction does not 
contribute, implying that standard matter-dominated solutions with $a^{-3}$ behavior can 
still appear. On the other hand, when  dark energy becomes relevant the evolution of dark 
matter changes and deviates from the $a^{-3}$ behavior. This feature explains why scaling 
solutions with effective equation-of-state parameter different from $w$ are possible, 
which is again a standard and well known result from interacting dark energy models for 
which accelerated scaling solutions can be obtained (see Sec.~6 in \cite{DSreview}).

In summary, in the scenario considered in this work, we have the following
sectors: the scalar-field sector which is responsible for dark energy, the
mimetic-matter sector which is responsible for dark matter, and the usual standard-model
matter which can be either dust matter (in the case of $w=0$) or radiation (in the case
of
$w=1/3$). The various density parameters are defined as usual as
\begin{eqnarray}
&&\Omega_{de}= \Omega_\phi\equiv\frac{\kappa^2\,V}{3H^2}
\label{Omphi}\,,\\
&&\Omega_{dm}=  \Omega_{\lambda}\equiv\frac{\kappa^2\lambda}{3 H^2}
\label{Omdarkmatter}\,,\\
 &&\Omega_m\equiv\frac{\kappa^2\rho}{3 H^2}.
  \label{Ommatter}
\end{eqnarray}
Furthermore, we can define the effective (total) energy density and pressure of the
universe as
\begin{align}
\rho_{\rm eff}&=\rho+\lambda+V,\\
p_{\rm eff}&=p-V,
\end{align}
and then the effective equation-of-state parameter as
\begin{align}
w_{\rm eff}=&\frac{p-V}{\rho+\lambda+V},
  \label{weffdef}
\end{align}
which is a very useful quantity since it is straightforwardly related to the deceleration
parameter $q$ through
\begin{equation}
q\equiv -1-\frac{\dot{H}}{H^2}=\frac{1+3 w_{\rm eff}}{2}.
\label{deccelmodI}
\end{equation}
Thus, $w_{\rm eff}<-1/3$ implies that the universe is accelerating.

\section{Dynamical system analysis}
\label{sec:dynamical_system_analysis}

In the previous section we presented the cosmological equations of mimetic gravity, in
the case of a flat FRW geometry. In this section we are interested in performing the
full phase-space analysis, for which one usually introduces suitable dimensionless
variables in
order to re-write the cosmological equations as an
autonomous dynamical system \citep{coley:2003mj,Leon2011,boehmer:2014vea,DSreview}.
Hence, in order to transform
\eqref{eq:cosmo_1}--\eqref{eq:cosmo_3}
into an autonomous system we define
\begin{equation}
    \sigma = \frac{\kappa \sqrt{\rho}}{\sqrt{3} H} \,, \quad x = \frac{\kappa^2
\lambda}{3H^2} \,, \quad y = \frac{\kappa \sqrt{V}}{\sqrt{3} H} \,, \quad z =
-\frac{1}{\kappa V^{3/2}}\frac{d V}{d\phi}
 \,.
    \label{eq:DS_variables}
\end{equation}
Note that $y$ and $\sigma$ cannot be negative, while the sign of $x$ depends on the sign
of $\lambda$.
Using these variables the Friedmann equation \eqref{eq:cosmo_1} yields the simple
constraint
\begin{equation}
    1 = x + y^2 + \sigma^2 \,,
    \label{eq:constraint}
\end{equation}
which can in fact be used to replace $\sigma$ in terms of $x$ and $y$ in the equations
that follow, reducing the dimensionality of the autonomous system. Using these auxiliary
variables the equations become
\begin{align}
    x' &= -3 w x^2-3 x \left[(w+1) y^2-w\right]+\sqrt{3} y^3 z \,, \label{eq:x} \\
    y' &= -\frac{1}{2} y \left[3 w x+3 (w+1) y^2-3 (w+1)+\sqrt{3} y z\right] \,,
\label{eq:y} \\
    z' &= -\sqrt{3} y z^2 \left( \Gamma - \frac{3}{2} \right) \,, \label{eq:z}
\end{align}
where primes denote differentiation with respect to $N=\ln a$. Moreover, we have defined
\begin{equation}
    \Gamma = \frac{V \ddot{V}}{\dot{V}^2} =\frac{V V_{\phi\phi}}{V_\phi^2} \,,
    \label{Gammadef}
\end{equation}
where the subscript $\phi$ denotes differentiation with respect to $\phi$, and the last
equality arises from  \eqref{eq:phi_eq_t}, namely from the fact that $\phi=t$. Lastly,
the three
dimensional
phase
space of the system \eqref{eq:x}-\eqref{eq:z} is given by
\begin{align}
\Psi =\left\lbrace (x,y) \in \mathbb{R}^2\, \vert\, - \infty < x \leq 1-y^2, -\infty <y<
\infty  \right\rbrace \times \left\lbrace z \in \mathbb{R} \right\rbrace.
\end{align}

In terms of the auxiliary variables, the various density parameters
(\ref{Omphi})-(\ref{Ommatter})
can be expressed as
\begin{equation}
  \Omega_m= 1-x-y^2 \,, \quad
  \Omega_\phi=y^2 \,, \quad
  \Omega_{\lambda}=x \,,
  \label{densityparamexyz}
\end{equation}
while the effective  equation-of-state parameter (\ref{weffdef}) becomes
\begin{align}
w_{\rm eff}=& w (1-x-y^2)-y^2,
\label{weffxyz}
\end{align}
with $w=p/\rho$ the matter equation-of-state parameter.

In order to perform the dynamical analysis we first need to extract the critical points
of the system by setting the left hand side of equations \eqref{eq:x}-\eqref{eq:z} to
zero. Then we perturb the system around these critical points, and thus the type and
stability of each point is determined by the eigenvalues of the involved perturbation
matrix \citep{coley:2003mj,Leon2011,boehmer:2014vea,DSreview}. Since in our model we have
the presence of the potential $V(\phi)$, in the following subsections we analyze various
cases separately.

\subsection{Mimetic gravity with a general scalar field potential}
\label{sub:mimetic_gravity_with_a_scalar_field_potential}

We start our dynamical analysis keeping the potential general. Observing
the definition of $z$ in (\ref{eq:DS_variables}), as well as  (\ref{Gammadef}), a general
potential simply implies that
$\Gamma$ can be written as a function of $z$, namely $\Gamma(z)$.
This encompasses a large variety of scalar field potentials, including the examples
analyzed in the next subsections. In the following we use $z_*$ to denote the solution
of the equation $\Gamma(z)-\frac{3}{2}=0$ and $\Gamma_z(z)$ to denote the derivative
$d\Gamma(z)/dz$.
Note that whenever $y\neq 0$, the condition $z=0$ is not sufficient for making the $z'$
equation, namely Eq.~\eqref{eq:z}, vanish, since a specific scalar field potential
could actually induce the expression $z^2 \Gamma(z)\neq 0$ for $z=0$.
Hence the two general conditions for attaining $z'=0$ are either $y=0$ or $z^2
\left[\Gamma(z)-\frac{3}{2}\right]=0$.

The physical critical points and curves of critical points for the system of equations
(\ref{eq:x})-(\ref{eq:z}) for a general
scalar field potentials are given in Table \ref{tab:c_pts_general}, along with their
existence conditions. Additionally, in the same Table we have added the
corresponding values of the various density parameters from (\ref{densityparamexyz}), as
well as the value of the effective  equation-of-state parameter from (\ref{weffxyz}).
According to $y$-definition in (\ref{eq:DS_variables}), points with $y>0$ correspond to
$H>0$, i.e.~to expanding universes, while points with $y<0$ correspond to $H<0$, i.e.~to
contracting universes, and thus we respectively add the subscripts $+$ or $-$ to the
corresponding points. Furthermore, for each critical point in Table
\ref{tab:eigen_general} we present the eigenvalues of its involved  Jacobian
(perturbation) matrix, and the corresponding stability conditions.
\begin{table}[!ht]
\centering
\resizebox{\columnwidth}{!}{
\begin{tabular}{ccccccccc}
  \hline\hline
  Point & $x$ &     $y$ &   $z$ &Existence      & $\Omega_\phi$  &
$\Omega_\lambda$ & $\Omega_m$  & $w_{\rm eff}$   \\
  \hline
 $A_{1}$ &     $1$    &     0      &      $z$   &Always &  0   &    1 &  0  &    $0$
\\[1.5ex]
  $A_2$   &    $0$     &   $0$     &      $z$    &Always  &  0   &  0 &  1  & $w$
\\[1.5ex]
$A_{3+}$ &   $0$    &   $1$   &     $0$    & $z^2 \Gamma(z)\!=\!0$ for $z\!=\!0$
&  1
&  0 &  0  &     $-1$
   \\[1.5ex]
   $A_{3-}$ &   $0$    &   $- 1$   &     $0$    & $z^2 \Gamma(z)\!=\!0$ for $z\!=\!0$
 &  1
&  0 &  0  &     $-1$
   \\[1.5ex]
  $A_4$   &    $-\frac{3(1+w)^3}{wz_*^2}$     &   $\frac{\sqrt{3}(w+1)}{z_*}$     &
$z_*$
&
 $z_* \neq 0$,  $w\neq 0$   & $ \frac{3(w+1)^2}{z_*^2}$   &
$-\frac{3(1+w)^3}{wz_*^2}$   &  $1\!+\!\frac{3(w+1)^2}{w z_*^2}   $ &
$w$   \\[1.5ex]
   $A_{5+}$   &   $1-\frac{y_{+}^2}{12}$     &  $\frac{\sqrt{3}y_{+}}{6}$  &
$z_*$    &
Always  &  $\frac{y_{+}^2}{12}$   &  $1-\frac{y_{+}^2}{12}$ &  0   &
$-\frac{y_{+}^2}{12}$
\\[1.5ex]
 $A_{5-}$   &   $1-\frac{y_{-}^2}{12}$     &  $\frac{\sqrt{3}y_{-}}{6}$  &
$z_*$    &
Always  &  $\frac{y_{-}^2}{12}$   &  $1-\frac{y_{-}^2}{12}$ &  0   &
$-\frac{y_{-}^2}{12}$
\\[1.5ex]
  \hline\hline
\end{tabular}}
\caption{The physical critical points and curves of critical points of the system
(\ref{eq:x})-(\ref{eq:z}),  for the case of general potential, and their existence
conditions. Additionally, we present the corresponding values of the various density
parameters from (\ref{densityparamexyz}), as well as the value of the effective
equation-of-state parameter from (\ref{weffxyz}).
We have defined $y_{\pm}=-z_*\pm \sqrt{z_*^2+12}$, with $z_*$ denoting the solution of
the equation $\Gamma(z)-\frac{3}{2}=0$.
}
\label{tab:c_pts_general}
\end{table}
\begin{table}
\centering
\resizebox{\columnwidth}{!}{
\begin{tabular}{ccccc}
  \hline\hline
  Point & $\lambda_1$ & $\lambda_2$ & $\lambda_3$ & Stability\\
  \hline
 $A_{1}$ &     $0$    &     $\frac{3}{2}$      &  $-3w$ & non-hyperbolic, behaves as
saddle point for $w\neq0$     \\[1.5ex]
 &        &          &   &   behaves as unstable node for $w=0$     \\[1.5ex]
  $A_2$   &    $0$     &  $\frac{3}{2}(1+w)$       &      $3w$   &non-hyperbolic, behaves
as unstable node   \\[1.5ex]
$A_{3+}$ &   $0$    &   $-3$   &     $-3(1+w)$      &non-hyperbolic,  behaves as saddle
point for   $\Gamma(0)\neq\frac{3}{2}$
  \\[1.5ex]
 &       &      &           &
  behaves as stable
for    $\Gamma(0)=\frac{3}{2}$    and  $\Gamma_z(0)>0$   \\[1.5ex]
$A_{3-}$ &   $0$    &   $-3$   &     $-3(1+w)$      &non-hyperbolic, behaves as saddle
point
for    $\Gamma(0)\neq\frac{3}{2}$
\\[1.5ex]
 &       &      &           &
  behaves as stable
for    $\Gamma(0)=\frac{3}{2}$    and  $\Gamma_z(0)<0$   \\[1.5ex]
 $A_4$   &      $\eta_{+}$     &   $\eta_{-}$       &      $-3(w+1)\, z_*\, \Gamma_z(z_*)
   $      & saddle point
  \\[1.5ex]
  $A_{5+}$   &      $\mu_{+}$     &   $\mu_{-}$       &      $-\frac{z_*^2}{2}
\left(\sqrt{z_*^2+12}
-z_*\right) \, \Gamma_z(z_*)$   &stable for  $\Gamma_z(z_*)>0$
\\[1.5ex]
   $A_{5-}$   &      $\nu_{+}$     &   $\nu_{-}$       &      $\frac{z_*^2}{2}
\left(\sqrt{z_*^2+12}
+z_*\right) \, \Gamma_z(z_*)$    &stable  for $\Gamma_z(z_*)<0$
\\[1.5ex]
\hline\hline
\end{tabular}}
\caption{The eigenvalues of the perturbation matrix and the implied stability conditions,
for the critical points and curves of critical points of the system
(\ref{eq:x})-(\ref{eq:z}), for the case of general potential.
We have defined
 $y_{\pm}=-z_*\pm \sqrt{z_*^2+12}$,  with $z_*$ denoting the solution of the
equation $\Gamma(z)-\frac{3}{2}=0$ and with $\Gamma_z(z)$ denoting $d\Gamma(z)/dz$.
Additionally, we have defined  $\eta_{\pm}=\frac{3}{4} \left[(w-1)\pm\,\sqrt {
\frac{24(1+w)^3}{z_*^2}+ \left( 3\,w+1 \right)
^{2}}\right]$,
  $\mu_{\pm}=\frac{3}{8}\,z_*\,y_{+}\!-\!\frac{3}{2}(w+6)\! \pm\!
\frac{\sqrt{2}}{8}\sqrt{(z_*^2+12w)(6-z_*y_{+})-72w(1-w) }$, and
  $\nu_{\pm}=\frac{3}{8}\,z_*\,y_{-}\!-\!\frac{3}{2}(w+6)\! \pm\!
\frac{\sqrt{2}}{8}\sqrt{(z_*^2+12w)(6-z_*y_{-})-72w(1-w) }$.}
\label{tab:eigen_general}
\end{table}

Let us summarize the dynamical analysis results for this general potential case.

\begin{figure}[ht]
\centering
\subfigure[]{%
\includegraphics[width=6.5cm]{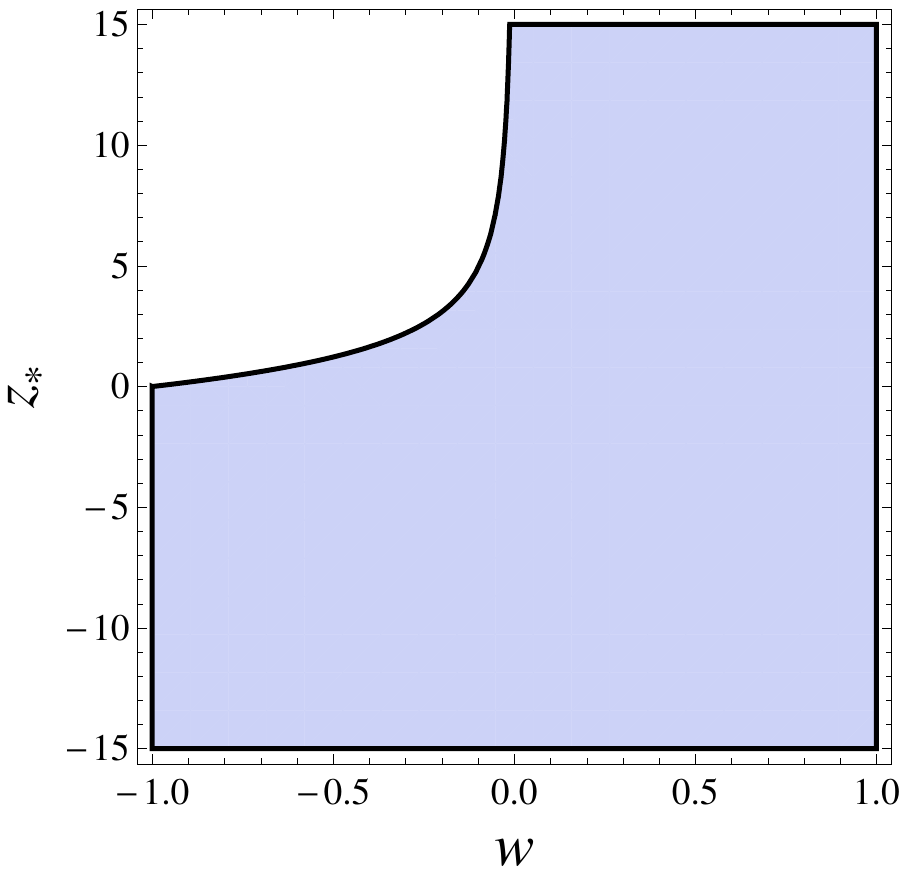}\label{fig:rgn_plot_a5p}}
\subfigure[]{%
\includegraphics[width=6.5cm]{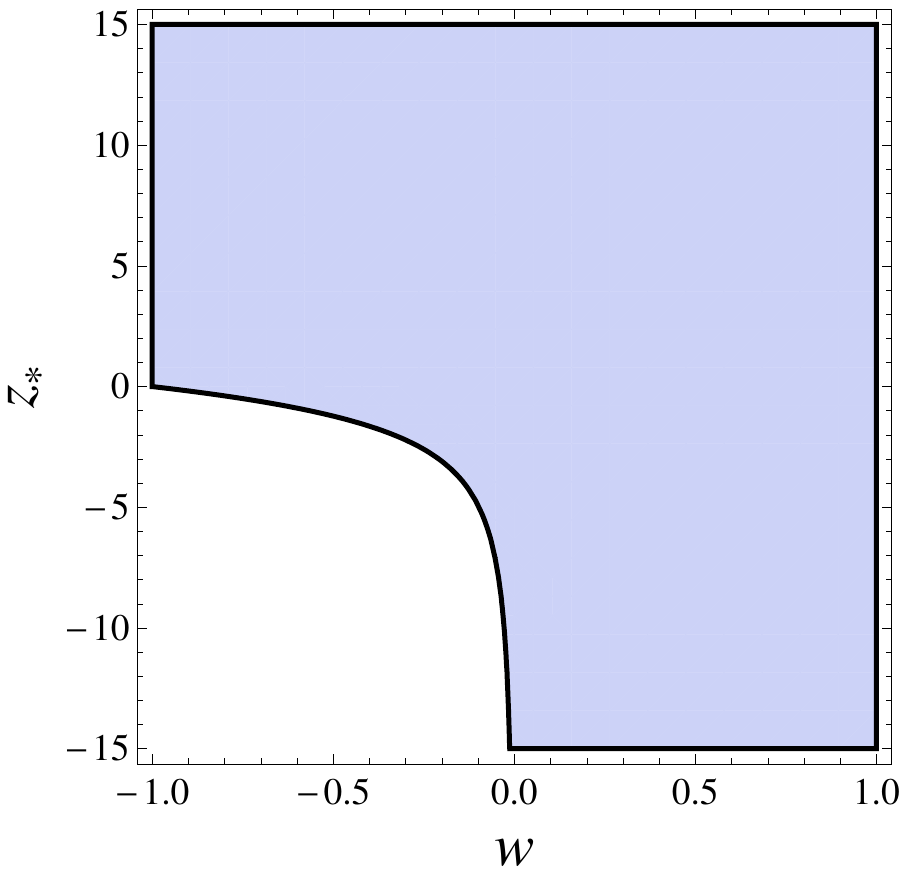}\label{fig:rgn_plot_a5n}}
\caption{{\it{Left graph: The shaded regions mark the stability region of point
$A_{5+}$ in the $(w,z_*)$ plane,  when $\Gamma_z(z_*)>0$.
Right graph: The stability region of point
$A_{5-}$ in the $(w,z_*)$ plane,  when $\Gamma_z(z_*)<0$. }}}
\label{fig:region_sca_pert}
\end{figure}

\begin{itemize}
\item  $A_1$ is a curve of critical points (each one is obtained by a different value of
$z$) which always exists, independently of the potential choice. In these points we have
$\Omega_\lambda=1$, and thus $A_1$  corresponds to a mimetic matter dominated universe,
with effective equation of state $w_{\rm eff}=0$ and thus non-accelerating. Since the
points of this critical curve behave in general as saddle (see Appendix
\ref{A1A2stability}) they
cannot attract the universe at late times, however the universe
can remain in this state for a long period at intermediate times. Thus, critical curve
$A_1$ can very efficiently describe the transient matter-dominated era of the universe
history.

\item $A_2$ is a curve of critical points, it always exists, independently of the
potential choice, and in these states we have $\Omega_{m}=1$. Hence, they correspond to a
universe dominated by the standard-model matter sector, i.e.~by dust matter in the case
of
$w=0$ or by radiation in the case where $w=1/3$. Since $w_{\rm eff}=w$, in both cases the
universe is non-accelerating. Finally, these points behave as unstable nodes (see
Appendix
\ref{A1A2stability}).

\item Critical point $A_{3+}$ exists for $z=0$ and $z^2\,\Gamma(z)=0$, condition which
depends heavily on the potential choice (note that since $z=0$ the condition $z^2
\left[\Gamma(z)-\frac{3}{2}\right]=0$ reduces to $z^2 \Gamma(z)=0$). It corresponds to a
scalar-field (i.e.~dark-energy) dominated, expanding de Sitter universe. Since it is
non-hyperbolic, with all non
vanishing eigenvalues having negative real part, its stability must be determined using
the center manifold methods \cite{Leon2011}. The relevant analysis of the dynamics near
the center
manifold is provided in Appendix \ref{CMTA3} and in Table \ref{tab:eigen_general} we
summarize
the results. In particular, point $A_{3+}$ behaves as saddle for potentials where
$\Gamma(0) \neq \frac{3}{2}$, while in the cases where $\Gamma(0)=\frac{3}{2}$ it behaves
as stable when $\Gamma_z(0)>0$. The fact that it is dark-energy dominated and stable
makes this point a very good candidate for the description of late-time universe.

\item Critical point $A_{3-}$ is the contracting counterpart of  $A_{3+}$. It corresponds
to a scalar-field dominated, contracted universe, which  behaves as saddle for potentials
where $\Gamma(0) \neq \frac{3}{2}$, and as stable in the cases where
$\Gamma(0)=\frac{3}{2}$ with $\Gamma_z(0)<0$.

\item Critical point $A_4$ exists for $z_* \neq 0$ and $w\neq 0$, and its features depend
on
the specific potential form. It has the interesting property that it can alleviate the
cosmic
coincidence problem, since in these scaling solutions the dark-energy and dark-matter
density
parameters can be comparable in magnitude.
However since $w_{\rm eff}=w$ the universe is non-accelerating for radiation or dust
matter fluids.
Additionally, since $\eta_+>0$ and $\eta_-<0$ its eigenvalues are
always of different sign and thus this point behaves always as saddle.

\item  Critical point $A_{5+}$ exists always, its features depend on the specific
potential form, and it corresponds to an expanding universe (actually point  $A_{3+}$ is
a
special case of  $A_{5+}$, namely for potentials that have $z_*=0$). It can alleviate the
coincidence problem, since the dark-energy and dark-matter density parameters can be of
the
same order, and furthermore for particular values of $z_* $, i.e.~of
the potential form, the universe can be accelerating. Since the eigenvalues  $\mu_{\pm}$
for $0\leq w\leq1$ are always negative (as can been easily confirmed), its stability
depends upon the signature of the eigenvalue $\lambda_3$, and thus point $A_{5+}$ is
stable for potentials with $\Gamma_z(z_*)>0$. For the general case $-1\leq w\leq1$, and
with $\Gamma_z(z_*)>0$, the stability region of point $A_{5+}$  in the $(w,z_*)$ plane is
depicted in the left graph of Fig.~\ref{fig:region_sca_pert}. The fact that this point
can
be stable and with features in agreement with observations, makes it
a very good candidate for the description of the late-time universe.

\item  Critical point $A_{5-}$ is the contracting counterpart of  $A_{5+}$ (again,
$A_{3-}$ is a special case of  $A_{5-}$, for potentials that have $z_*=0$). For
$0\leq w\leq1$ it is stable when $\Gamma_z(z_*)<0$, while for the general case $-1\leq
w\leq1$, and with $\Gamma_z(z_*)<0$, its stability region  in the
$(w,z_*)$ plane is depicted in the right graph of Fig.~\ref{fig:region_sca_pert}.

\end{itemize}

The above general dynamical analysis reveals that, irrespectively of the form of the
scalar field potential, the cosmological behavior of mimetic gravity will always admit
both matter dominated solutions (curves $A_1$ and $A_2$) and dark-energy dominated
solutions mimicking a cosmological constant behavior (points~$A_{3+}$). Moreover, the
phase space presents scaling solutions (points~$A_4$ and $A_{5+}$) too, in which the
matter
and dark-energy density parameters are of the same order and thus they can offer an
alleviation to the coincidence problem, where the exact behavior of the accelerating
expansion
depends on the specific
potential form.

In order to proceed from the above general examination  to a more precise analysis we
need to specify the potential $V(\phi)$. In the following three subsections we
investigate
some well-known scalar field potentials separately.

\subsection{Mimetic gravity with an inverse square potential}
\label{sub:mimetic_gravity_with_an_inverse_square_potential}

Let us start with the study of the inverse square scalar-field potential, namely we
consider
\begin{equation}
    V(\phi) = \alpha \phi^{-2} \,,
    \label{eq:inv_sqr_pot}
\end{equation}
with $\alpha>0$ the potential parameter.
This potential form has the special property that, through (\ref{Gammadef}), it yields
$\Gamma = 3/2$ and consequently $z' = 0$ from Eq.~\eqref{eq:z}.
This implies that $z_*$ in Tab.~\ref{tab:c_pts_general} can be treated as an arbitrary
constant.
From the
definition of $z$ in Eq.~\eqref{eq:DS_variables}, we deduce that
\begin{equation}
  z = \frac{2}{\kappa \sqrt{\alpha}} \,,
\end{equation}
and thus $z_*$ is actually related to $\alpha$.
The fact that the variable $z$ has a
constant value implies that in this special potential case  the autonomous system
\eqref{eq:x}--\eqref{eq:z} becomes two-dimensional.

The critical points of the system and their properties for this potential case can be
extracted from the general Tables  \ref{tab:c_pts_general} and
\ref{tab:eigen_general}, through the replacement $z_*= \frac{2}{\kappa \sqrt{\alpha}}$.
Hence, in Table \ref{tab:c_pts_inv_sq} we present the physical critical points of
the system (\ref{eq:x})-(\ref{eq:z}), the corresponding values of the various density
parameters from (\ref{densityparamexyz}), as well as the value of the effective
equation-of-state parameter from (\ref{weffxyz}). Additionally, in Table
\ref{tab:c_pts_inv_sq2} we list the eigenvalues of the
perturbation matrix and the implied stability conditions.
 \begin{table}[!ht]
\centering
\resizebox{\columnwidth}{!}{
\begin{tabular}{cccccccc}
  \hline\hline
  Point & $x$ &     $y$   &Existence      & $\Omega_\phi$  &
$\Omega_\lambda$ & $\Omega_m$  & $w_{\rm eff}$
  \\
  \hline
 $A_{1}$ &     $1$    &     0        &Always &  0   &    1 &  0  &    $0$
\\[1.5ex]
  $A_2$   &    $0$     &   $0$        &Always  &  0   &  0 &  1  & $w$
\\[1.5ex]
  $A_4$   &    $-\frac{3\kappa^2\alpha (1+w)^3}{4w}$     &
$\frac{\sqrt{3\alpha}\,\kappa\,(w+1)}{2}$     &
   $w\neq 0$   &   $\frac{3\kappa^2\alpha (1+w)^2}{4}$    &
 $-\frac{3\kappa^2\alpha (1+w)^3}{4w}$    &  $1\!+\! \frac{3\kappa^2\alpha (1+w)^2}{4w}
$ &
$w$   \\[1.5ex]
   $A_{5+}$   &   $1-\frac{y_{+}^2}{12}$     &  $\frac{\sqrt{3}y_{+}}{6}$  &
 $\alpha\neq0$  &  $\frac{y_{+}^2}{12}$   &  $1-\frac{y_{+}^2}{12}$ &  0   &
$-\frac{y_{+}^2}{12}$
\\[1.5ex]
 $A_{5-}$   &   $1-\frac{y_{-}^2}{12}$     &  $\frac{\sqrt{3}y_{-}}{6}$  &
 $\alpha\neq0$  &  $\frac{y_{-}^2}{12}$   &  $1-\frac{y_{-}^2}{12}$ &  0   &
$-\frac{y_{-}^2}{12}$
\\[1.5ex]
  \hline\hline
\end{tabular}}
\caption{
The physical critical points of the system (\ref{eq:x})-(\ref{eq:z}), for the
case of inverse square scalar-field potential $V=\alpha \phi^{-2}$, and their
existence conditions. Additionally, we present the corresponding values of the various
density parameters from (\ref{densityparamexyz}), as well as the value of the effective
equation-of-state parameter from (\ref{weffxyz}). We have defined
 $y_{\pm}=-\frac{2}{\kappa\sqrt{\alpha}}\pm \sqrt{\frac{4}{\kappa^2\alpha}+12}$.
}
\label{tab:c_pts_inv_sq}
\end{table}
  \begin{table}[!ht]
\centering
\begin{tabular}{cccc}
  \hline\hline
  Point   & $\lambda_1$ & $\lambda_2$ & Stability
  \\
  \hline
 $A_{1}$ &
$\frac{3}{2}$      &  $-3w$  & saddle
\\[1.5ex]
  $A_2$   &
$\frac{3}{2}(1+w)$       &      $3w$     & unstable
\\[1.5ex]
  $A_4$     &  $\eta_{+}$     &   $\eta_{-}$ &saddle \\[1.5ex]
   $A_{5+}$   &      $\mu_{+}$     &   $\mu_{-}$   & stable for orbits with
$y>0$
\\[1.5ex]
 $A_{5-}$   &        $\nu_{+}$     &
$\nu_{-}$ & stable for orbits with
$y<0$
\\[1.5ex]
  \hline\hline
\end{tabular}
\caption{
The eigenvalues of the perturbation matrix and the implied stability conditions,
for the critical points  of the system
(\ref{eq:x})-(\ref{eq:z}),  for the
case of inverse square scalar-field potential $V=\alpha \phi^{-2}$.    We have defined
 $y_{\pm}=-\frac{2}{\kappa\sqrt{\alpha}}\pm \sqrt{\frac{4}{\kappa^2\alpha}+12}$,
 $\eta_{\pm}=\frac{3}{4} \left[(w-1)\pm\,\sqrt {   12\kappa^2 \alpha (1+w)^3 
+ \left( 3\,w+1 \right)
^{2}}\right]$,
  $\mu_{\pm}=\frac{3}{4}\,\frac{y_{+}}{\kappa\sqrt{\alpha}}-\frac{3}{2}(w+6) \pm
\frac{1}{2}\sqrt{(\frac{1}{\kappa^2\alpha}+3w)(3-\frac{y_{+}}{\kappa\sqrt{\alpha}}
)-9w(1-w) }$,
  $\nu_{\pm}=\frac{3}{4}\,\frac{y_{-}}{\kappa\sqrt{\alpha}}-\frac{3}{2}(w+6) \pm
\frac{1}{2}\sqrt{(\frac{1}{\kappa^2\alpha}+3w)(3-\frac{y_{-}}{\kappa\sqrt{\alpha}}
)-9w(1-w) }$.
}
\label{tab:c_pts_inv_sq2}
\end{table}

The critical curves $A_1$ and $A_2$ of the general case have now become individual
critical points. Point $A_1$  corresponds to a mimetic matter dominated, non-accelerating
universe, which is a saddle and thus can describe the matter-dominated era of the
universe
history at intermediate times. Point $A_2$ corresponds to a non-accelerating
universe dominated by the standard-model matter sector (by dust matter in the case of
$w=0$ or by radiation in the case where $w=1/3$), and is unstable.
Moreover, the critical points $A_{3\pm}$ of the general case do not exist for this
specific potential, since they require $z=0$. Furthermore, critical point $A_4$ exists
for $w\neq 0$, it is always saddle, and it corresponds to a non-accelerating universe
with
scaling behavior, which can alleviate the coincidence problem.

Critical point $A_{5+}$ corresponds to a universe with scaling behavior, and thus can
alleviate the coincidence problem. Additionally, since $w_{\rm eff}=-\frac{y_{+}^2}{12}$
we deduce that we obtain acceleration for $-y_{+}^2/12<-1/3$, i.e.~for
$\kappa^2\alpha>1$. Since the eigenvalues  $\mu_{\pm}$ for $0\leq w\leq1$ are always
negative (as can be easily confirmed), this point is always stable and therefore can
correspond the late-time evolution of the universe. It is the most interesting solution of
the
scenario at hand. Finally, point $A_{5-}$ is its contracting counterpart. Let us mention
here that for a given potential parameter both $A_{5+}$ and $A_{5-}$ are stable, however
since $y=0$ separates the phase space to two disconnected parts, orbits with
$y>0$ initially, i.e.~expanding universes, will always remain in the upper half of the
phase
space and thus they will be attracted by  $A_{5+}$ at late times, while orbits with $y<0$
initially, i.e.~contracting universes, will always remain in the lower part of the
phase space and thus they will be attracted by  $A_{5-}$.

\begin{figure}[!]
\centering
\subfigure[]{%
\includegraphics[width=7cm]{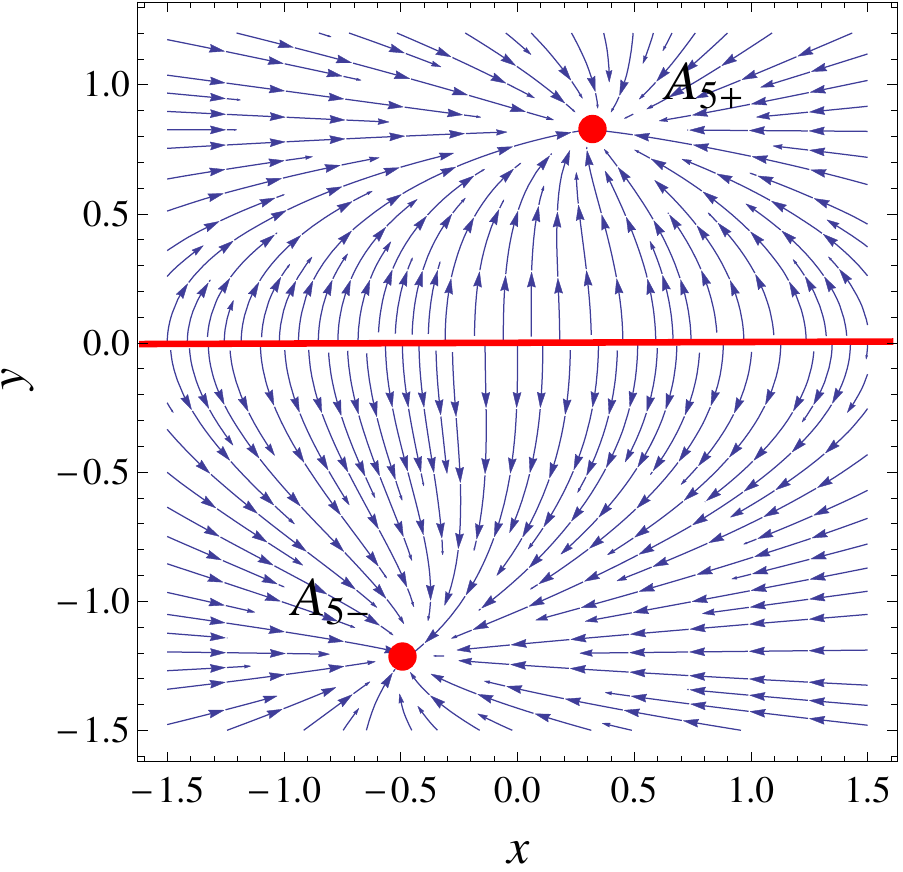}}
\quad
\subfigure[]{%
\includegraphics[width=7cm]{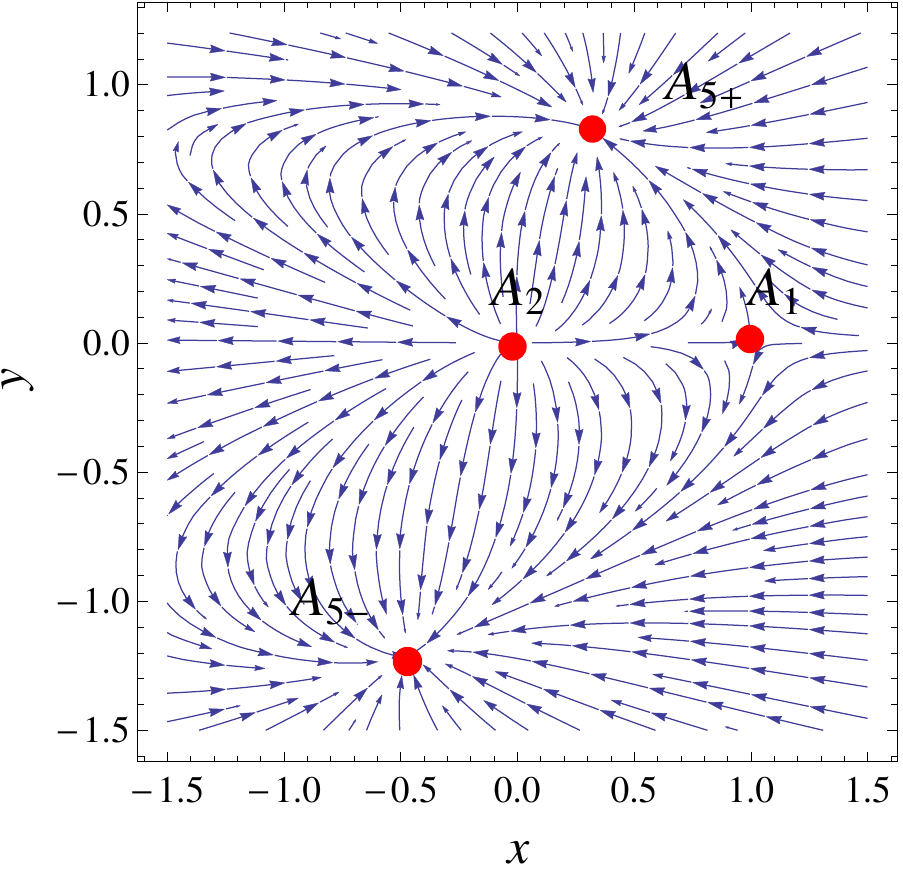}}
\caption{
{\it{ The phase-space behavior of the system \eqref{eq:x}-\eqref{eq:y}, for the
case of inverse square scalar-field potential $V=\alpha \phi^{-2}$, with the choice
$\alpha=9/\kappa^2$, for $w=0$ (left graph) and $w=1/3$ (right graph),
respectively. As we explain in the text, orbits with $y>0$ initially, i.e.~expanding
universes, will result to the (accelerating for this $\alpha$ value) scaling solution
$A_{5+}$, while orbits with $y<0$ initially, i.e.~contracting
universes, will result to the contracting counterpart $A_{5-}$.
}}
}
\label{fig:inv_sq_law_phse_trajctries}
\end{figure}

In order to present the above features in a more transparent way, we evolve the
autonomous
system numerically and in Fig.~\ref{fig:inv_sq_law_phse_trajctries} we depict the
resulting two-dimensional  phase-space behavior for two choices of $w$, namely for dust
matter with $w = 0$ (left graph) and for radiation with $w=1/3$ (right graph).

For the dust matter case of the left graph of
Fig.~\ref{fig:inv_sq_law_phse_trajctries}, we remind that $\Omega_m$ characterizes the
non-relativistic baryonic component, while $\Omega_\lambda$ constitutes the dark matter
sector and $\Omega_\phi$ the dark energy sector. Note that as we discuss in Appendix
\ref{A1A2stability}, for $w=0$ the whole $x$-axis
becomes a critical line. As we observe, we do verify
the theoretical prediction that  $y=0$ separates the phase space to two disconnected
parts, i.e.~to $y>0$, which corresponds to expanding universe, and to  $y<0$, which
corresponds to contracting universe. Hence, given any choice of parameters and
initial conditions, an expanding universe evolves from a matter dominated universe (line
$A_1$ or $A_2$) towards the scaling universe  $A_{5+}$, which for the specific choice
$\alpha=9/\kappa^2$ of the figure it is also accelerating. This cosmological behavior is
in agreement
with
observations. On the other hand, an initially contracting
universe will result to the contracting late-time attractor  $A_{5-}$.

For the radiation case depicted in the right graph of
Fig.~\ref{fig:inv_sq_law_phse_trajctries}, we remind that since $w=1/3$ then $\Omega_m$
characterizes the radiation component, while $\Omega_\lambda$ constitutes the dark matter
sector and $\Omega_\phi$ the dark energy sector. In this case, points $A_1$ or $A_2$ are
isolated critical points, with the first corresponding to dark-matter domination (since
$\Omega_\lambda=1$) while the second to radiation domination (since $\Omega_m=1$). This
scenario exhibits a very interesting behavior: the universe may start from the unstable
point $A_2$, come close to the saddle point $A_1$ and remain around it for sufficiently
long time, and finally result to the scaling (and accelerating for $\alpha=9/\kappa^2$)
universe  $A_{5+}$. Thus, we obtain the required thermal history of the universe, namely
the successive sequence of radiation, matter and acceleration eras.

\begin{figure}[!]
\centering
\subfigure[]{%
\includegraphics[width=7.4cm,height=5.5cm]{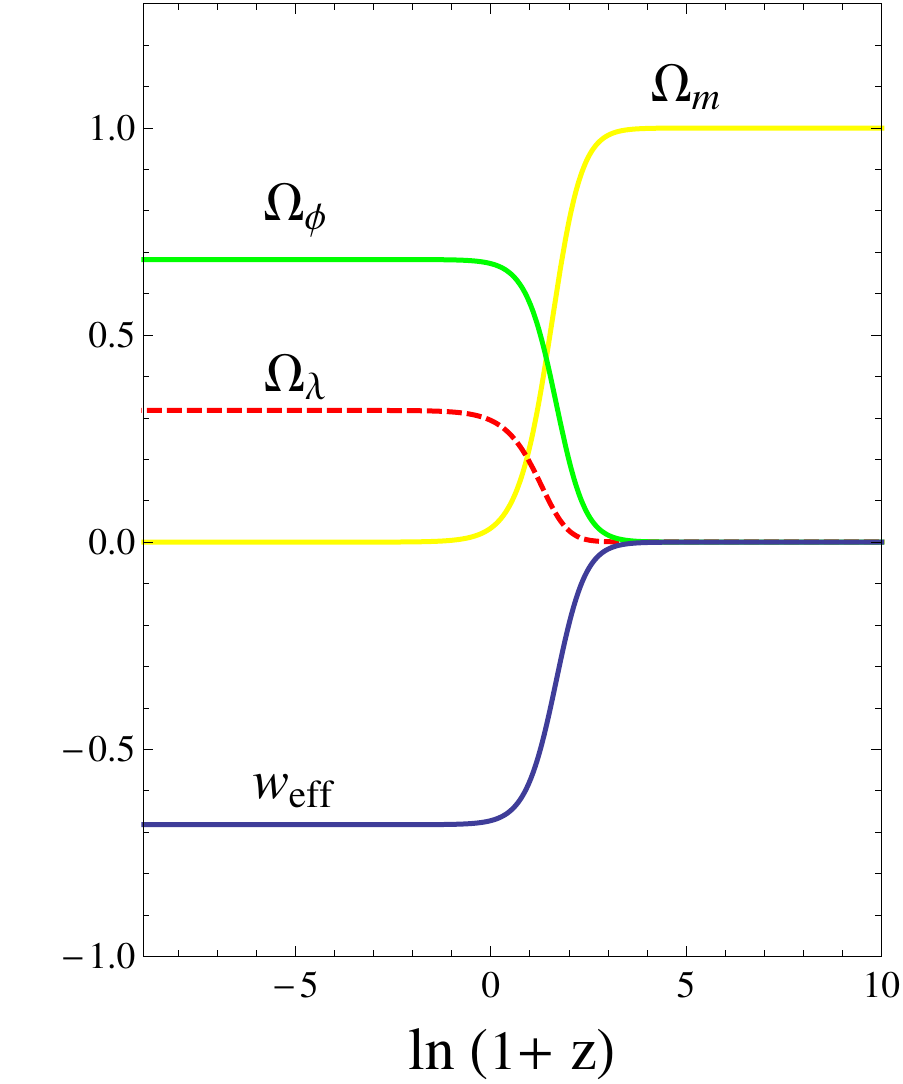}}
\quad
\subfigure[]{%
\includegraphics[width=7.4cm,height=5.5cm]{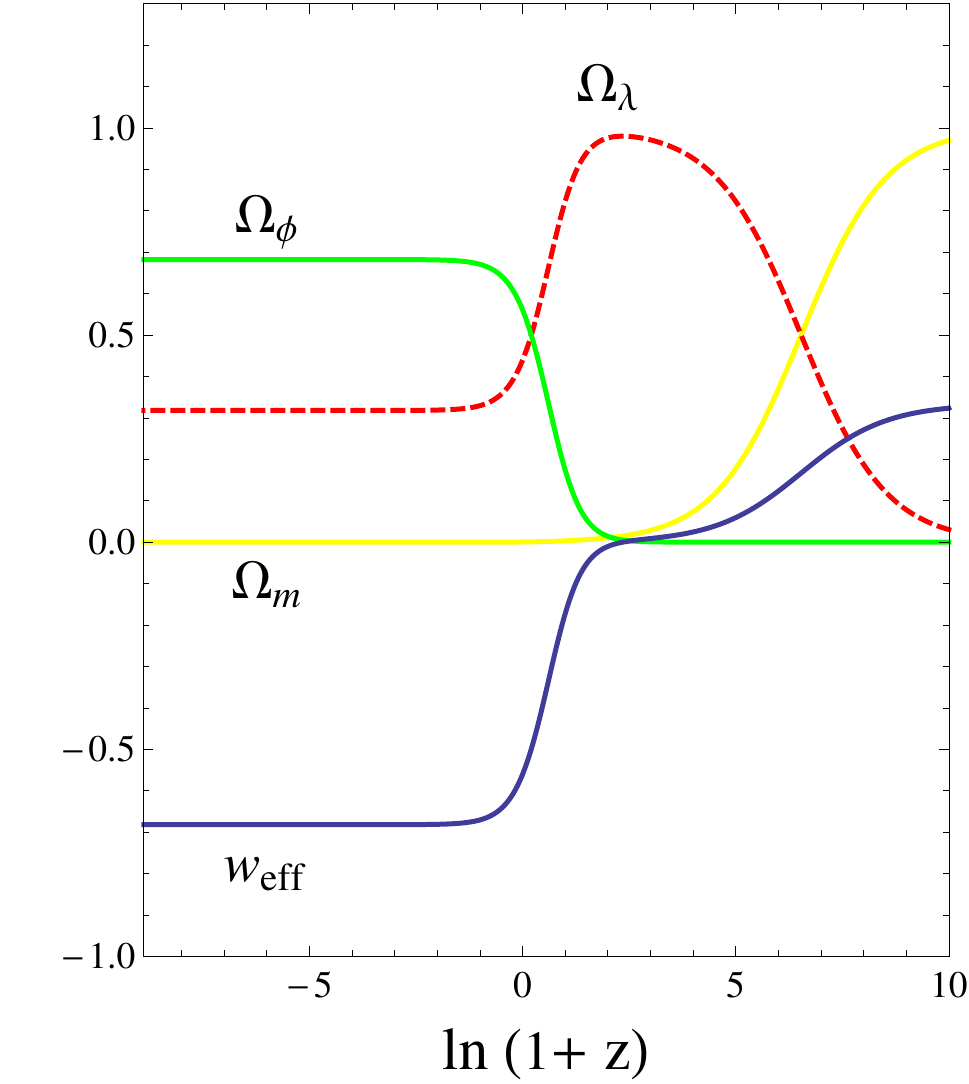}}
\caption{
{\it{
The evolution of the   density parameters  $\Omega_m$, $\Omega_{\phi}$,
$\Omega_\lambda$, as well as of the effective
equation-of-state parameter $w_{\rm eff}$ , as functions of the redshift, for the
case of inverse square scalar-field potential $V=\alpha \phi^{-2}$, with the choice
$\alpha=9/\kappa^2$, for $w=0$ (left graph) and $w=1/3$ (right graph),
respectively.
}}}
\label{fig:inv_sq_law_parameters}
\end{figure}

The above features are alternatively evident in Fig.~\ref{fig:inv_sq_law_parameters},
where we plot the evolutions of the various density parameters, as well as of the
effective
equation-of-state parameter $w_{\rm eff}$. For convenience, as independent variable we
use the redshift $z=a_0/a-1$ (with $a_0=1$ the present scale factor), and thus as
usual $z=0$ corresponds to the present time while $z\rightarrow-1$ corresponds to the
infinite
future.

For the dust case in the left graph of Fig.~\ref{fig:inv_sq_law_parameters} we do
observe the aforementioned behavior, namely
the universe transits from a matter to an acceleration era at late times, and
moreover the dark matter and dark energy density parameters remain, respectively, around
0.3 and 0.7 for ever after, therefore offering an
alleviation to the coincidence problem.

For the radiation case of the right  graph of Fig.~\ref{fig:inv_sq_law_parameters} we
can see that at early times the universe is radiation dominated, then it transits to
dark matter domination, and finally it results to dark energy domination. As before, the
universe results in a scaling accelerating solution, where the dark matter and dark
energy density parameters remain around 0.3 and 0.7 respectively for ever, offering an
alleviation to the coincidence problem. Furthermore, the early-time behavior is in
agreement with
observations, since the past attractor is described by a radiation dominated solution.
We stress that this is not the case for simpler scalar-field models
of dark energy, notably quintessence, since in those models the past attractor of the
system is always
a stiff-matter solution dominated by the kinetic energy
of the scalar field \citep{Copeland:1997et,Tamanini:2014mpa}.

In summary, as we can see, mimetic gravity with an inverse square potential can
describe very efficiently the expansion history of the universe, starting from early
radiation domination, transiting to the matter era at intermediate times, and resulting
to late-time acceleration, offering also an alleviation to the coincidence problem.

\subsection{Mimetic gravity with a power-law or exponential potential}
\label{sub:mimetic_gravity_with_a_power_law_or_exponential_potential}

In this subsection we investigate the cosmology of mimetic gravity with a power-law or an
exponential scalar-field potential, namely we consider
\begin{equation}
    V(\phi) \propto \phi^n \,,
    \label{eq:power_pot}
\end{equation}
with  $n \neq -2$ (the case $n=2$ was analyzed in the previous subsection),
or
\begin{equation}
   V(\phi) \propto e^{\alpha\phi} \,,
    \label{eq:exp_pot}
\end{equation}
with $\alpha$ the model parameter. For both these potential cases $\Gamma$ is constant,
and in particular $\Gamma = (n-1)/n$ for the power-law potential, while $\Gamma = 1$ for
the exponential potential. Hence in the following we perform a general
analysis of Eqs.~\eqref{eq:x}-\eqref{eq:z} considering
$\Gamma=const.\neq3/2$ (actually  $\Gamma =3/2$ only for the inverse square potential
case, and that is why we analyzed it separately in the previous subsection).

The critical points and their features will arise from the results of the general
potential case of Tables \ref{tab:c_pts_general} and
\ref{tab:eigen_general}, substituting $\Gamma=const.\neq3/2$. In this case the
equation $\Gamma(z)-\frac{3}{2}=0$ does not have any solution $z_*$, and thus critical
points $A_4$ and $A_{5\pm}$ do not exist.

Since the curves of critical points $A_1$ and $A_2$ are independent from the choice of
the potential, their cosmological properties that were discussed in the general potential
case of subsection \ref{sub:mimetic_gravity_with_a_scalar_field_potential} will hold here
too. In particular, saddle points $A_1$ correspond to a non-accelerating mimetic matter
dominated universe, while unstable points $A_2$  correspond to a non-accelerating
universe dominated by the standard-model matter sector, i.e.~by dust matter in the case
of
$w=0$ or by radiation in the case where $w=1/3$.

\begin{figure}[ht]
\centering
\includegraphics[width=8cm,height=5.5cm]{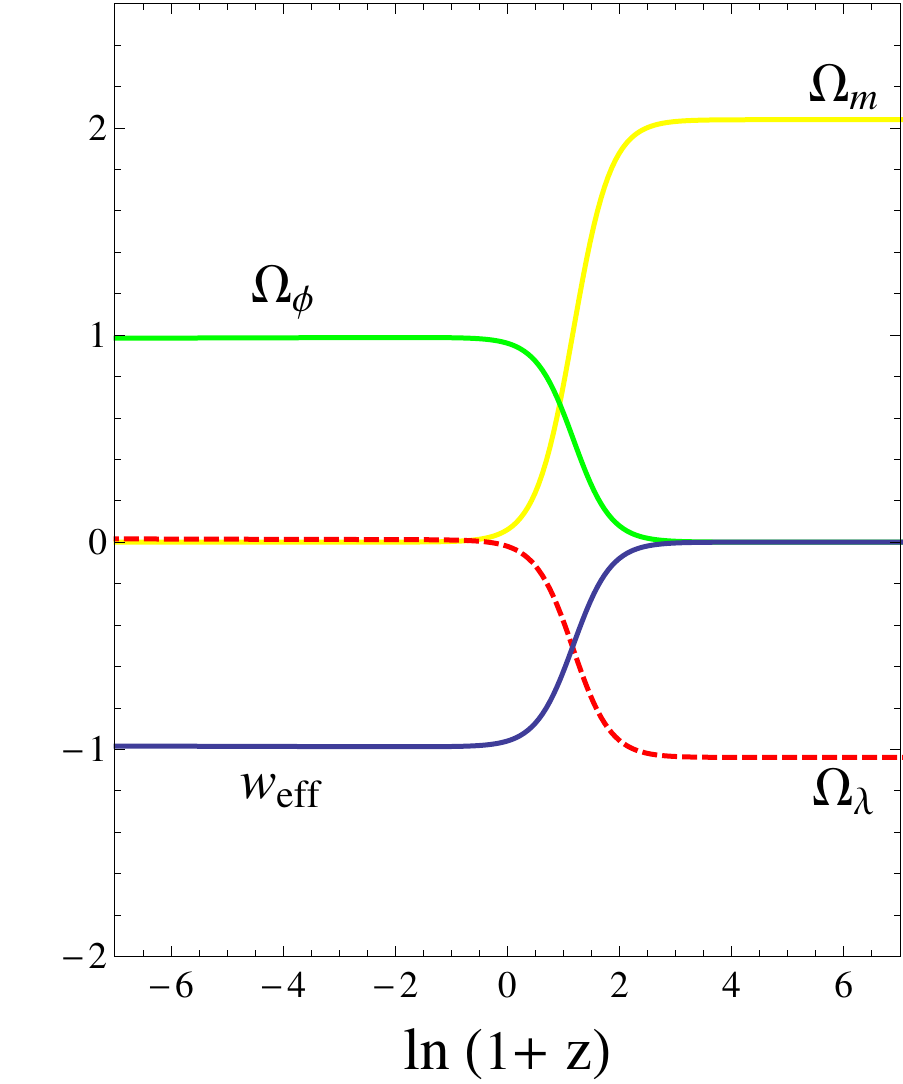}
\caption{{\it{
The evolution of the density parameters  $\Omega_m$, $\Omega_{\phi}$,
$\Omega_\lambda$, as well as of the effective
equation-of-state parameter $w_{\rm eff}$ , as functions of the redshift, for the
case of power-law scalar-field potential   $V(\phi) \propto \phi^n$, with the choice
$n=3$, for $w=0$.}}}
\label{fig:pow_law_parameters_3}
\end{figure}
\begin{figure}[ht]
\centering
\includegraphics[width=7cm]{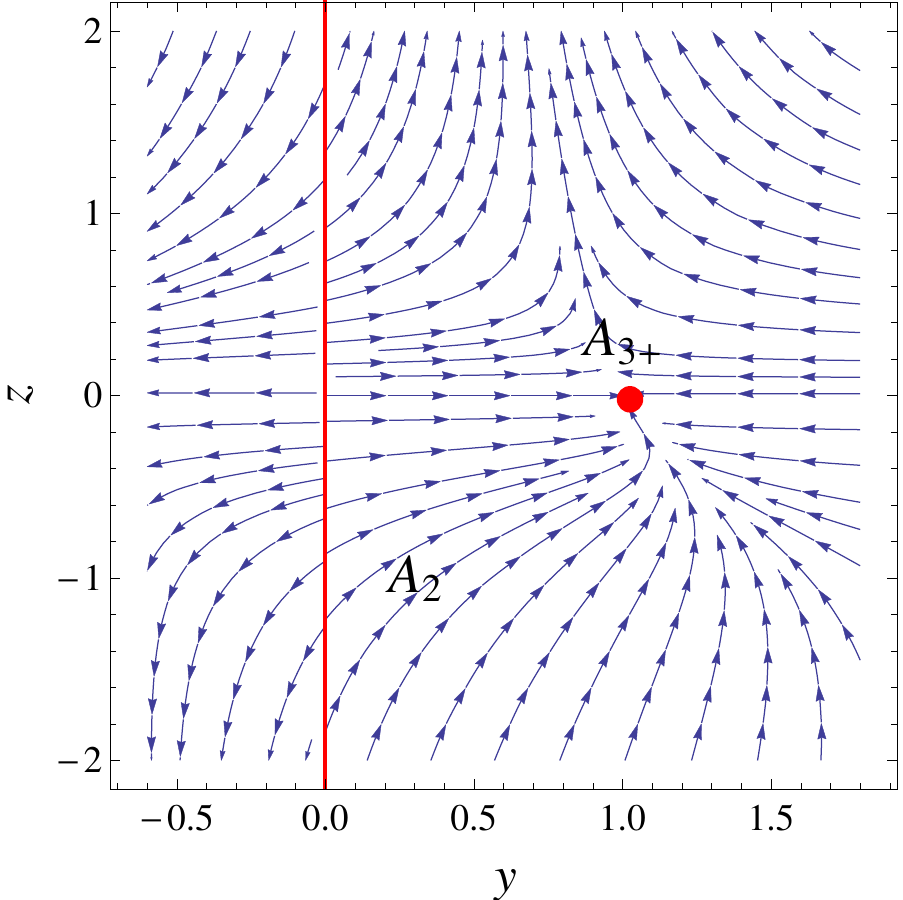}
\caption{
{\it{
The phase-space behavior of the system \eqref{eq:x}-\eqref{eq:y} projected on the $x=0$
plane, for the case of power-law scalar-field potential $V(\phi) \propto \phi^n$, with
the
choice $n=3$ and for $w=0$. The universe
starts from the matter-dominated critical line $A_2$, resulting in the dark-energy
dominated saddle de Sitter point $A_{3+}$. }}
}
\label{fig:pow_law_phse_trajctries}
\end{figure}
Furthermore, since now the system \eqref{eq:x}-\eqref{eq:z} is invariant under the
transformation $(y,z) \rightarrow (-y,-z)$, the stability features of points $A_{3+}$ and
$A_{3-}$ are the same. Since $\Gamma=const.\neq3/2$, from the analysis performed in
Sec.~\ref{sub:mimetic_gravity_with_a_scalar_field_potential}, and by employing center
manifold techniques, we conclude that in the present case both critical points $A_{3\pm}$
are saddle. $A_{3+}$  corresponds to a dark-energy dominated, expanding de Sitter
universe, while  $A_{3-}$ is its contracting counterpart.

Hence, for these particular choices of $V(\phi)$ there is no late time attractor in the
finite regime. The attractor exists at infinity, and its exact investigation requires
to apply the Poincar\'e central projection method \cite{Leon2011}. However, since this
analytical investigation lies beyond the scope of the present work, we examine the
cosmological behavior numerically. In Fig.~\ref{fig:pow_law_parameters_3} we depict the
evolutions of the various density parameters, as well as the effective
equation-of-state parameter $w_{\rm eff}$, as functions of the redshift. As we observe,
the universe starts evolving from a matter dominated phase, and then it transits to the
dark-energy dominated phase, resulting finally to the de Sitter point $A_{3+}$.
Nevertheless, since the de Sitter phase is not stable but saddle, the universe will
remain close to that for a finite time interval. In order to see this behavior more
transparently, in Fig.~\ref{fig:pow_law_phse_trajctries} we present the corresponding
phase-space behavior projected on the $x=0$ plane, where we also see that the universe
starts from the matter-dominated critical line $A_2$, resulting in the dark-energy
dominated de Sitter point $A_{3+}$.

Note that in Fig.~\ref{fig:pow_law_parameters_3} the relative energy densities
$\Omega_{m}$ and $\Omega_\lambda$ are not constrained in the interval $[-1,1]$, as
one
would expect. This feature usually arises in cosmological models where an interaction
between the matter component sourcing the cosmological equations is present (see
e.g.~\cite{Quartin:2008px}) and it is related to a certain ambiguity in defining the
relative energy densities that is present in this class of models \cite{Tamanini:2015iia}.
Since in our mimetic model the fields $\phi$ and $\lambda$ can be interpreted as
effectively interacting quantities (cf.~Eq.~\eqref{eq:cosmo_3}), then it is not
surprising
that the relative energy  densities $\Omega_{m}$ and $\Omega_\lambda$ can acquire
values outside the range $[-1,1]$.

\subsection{Mimetic gravity with $V(\phi) \propto (1+ \beta \phi^2)^{-2}$}
\label{sub:mimetic_gravity_with_buoncing_V}

In this subsection we consider the scalar potential  \cite{chamseddine:2014vna}
\begin{equation}
    V(\phi) = \alpha (1 + \beta \phi^2 )^{-2} \,,
    \label{eq:scal_pot_3}
\end{equation}
where $\alpha$ and $\beta$ are two parameters of suitable dimensions.
For this potential the variable $z$ defined in  (\ref{eq:DS_variables}) becomes
\begin{equation}
    z = \frac{4 \beta \phi}{\sqrt{\alpha} \kappa} \,,
\end{equation}
implying that $\Gamma$ in (\ref{Gammadef}) can be written   as
\begin{equation}
    \Gamma =   \frac{5}{4} \left(1 +
\frac{\xi}{5
z^2} \right) \,, \label{Gamma}
\end{equation}
where we have defined
\begin{equation}
    \xi = - \frac{16 \beta}{\alpha \kappa^2} \,.
    \label{eq:xi}
\end{equation}

The critical points and their features will arise from the results of the general
potential case of Tables \ref{tab:c_pts_general} and \ref{tab:eigen_general},
substituting
$\Gamma(z)$ from (\ref{Gamma}). In this case, Eq.~\eqref{eq:z} becomes
\begin{equation}\label{eq:zb}
    z' = \frac{\sqrt{3}}{4} y \left( z^2 - \xi \right) \,,
\end{equation}
and thus we have two solutions $z_*=\pm \sqrt{\xi}$. Hence, for this potential, points
$A_{3\pm}$ do not exist, while points $A_4$, $A_{5\pm}$ exist for $\xi>0$.
Critical point $A_4$ is saddle, since the eigenvalues $\eta_+$ and $\eta_-$ are of
opposite signs. Additionally, since there are two solutions $z_* = \pm \sqrt{\xi}$, there
are then two copies for each of the points $A_{5\pm}$. We denote these points as
$A^\pm_{5\pm}$, with the upper sign corresponding to one of the two solutions $z_* = \pm
\sqrt{\xi}$. For $z_*=\sqrt{\xi}$ critical point $A^+_{5-}$ is stable, but point
$A^+_{5+}$ is saddle  (as $\Gamma_z (\sqrt{\xi})<0$).
On the other hand, for $z_*=-\sqrt{\xi}$ point $A^-_{5+}$ is stable, while point
$A^-_{5-}$ is saddle (as $\Gamma_z (-\sqrt{\xi})>0$).
Finally, for $\xi<0$ only the critical lines $A_1$ and $A_2$ exist and hence there is no
finite late-time attractor in this case.

Point $A^-_{5+}$ corresponds to a phantom dark-energy dominated expanding solution, while
point $A^+_{5+}$ describes a scaling solution that can be accelerating.
In particular, the effective equation-of-state parameter for $A^-_{5+}$  always
satisfies $w_{\rm eff}<-1$, while for $A^+_{5+}$  it always satisfies $-1 < w_{\rm eff}
<0$, which provides accelerated expansion as long as $w_{\rm eff} < -1/3$.

\begin{figure}[ht]
\centering
\includegraphics[width=7cm]{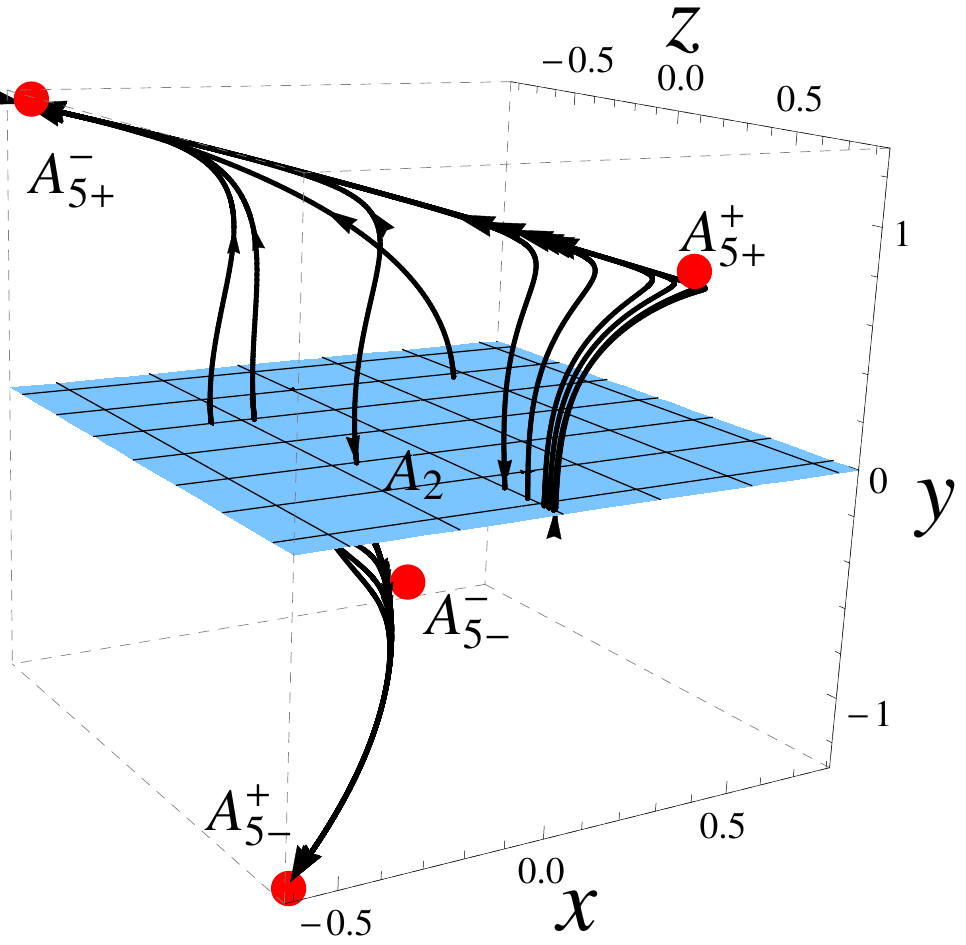}
\caption{{\it{Selected phase trajectories  of the system
\eqref{eq:x}-\eqref{eq:y} for the
potential $V(\phi) = \alpha (1 + \beta \phi^2 )^{-2}$, with $\xi= - \frac{16
\beta}{\alpha
\kappa^2}=0.3$ and for $w=0$. The expanding universe starts from a matter domination
solution
corresponding to the critical plane $A_2$, moving towards the accelerating phantom
scaling solution $A^-_{5+}$, and in some cases it also passes through the
intermediate long lasting accelerating scaling solution $A^+_{5+}$.  Additionally, one
may
also have a
similar behavior in the contracting counterparts, namely the contracting universe starts
from the matter dominated solution, evolving towards  $A^-_{5-}$, and in some case
passing also through the intermediate long lasting scaling solution
$A^+_{5-}$.
 }}
}
\label{fig:beta_pot_trajectories3D_A5n}
\end{figure}

In Fig.~\ref{fig:beta_pot_trajectories3D_A5n} we present the three-dimensional phase
space of the system \eqref{eq:x}-\eqref{eq:z}, for the potential \eqref{eq:scal_pot_3}
(with $\xi>0$).
Here the selected trajectories start evolving from a matter domination solution
corresponding to the critical plane $A_2$ towards the accelerating phantom scaling
solution $A^-_{5+}$, and in some cases they pass also through the intermediate long
lasting accelerating scaling solution $A^+_{5+}$. Additionally, one may also have a
similar behavior in the contracting counterparts, namely the contracting universe starts
from the matter dominated solution, evolving towards  $A^-_{5-}$, and in some case
passing also through the intermediate long lasting scaling solution
$A^+_{5-}$.

\begin{figure}[ht]
\centering
\subfigure[]{%
\includegraphics[width=7cm,height=5cm]{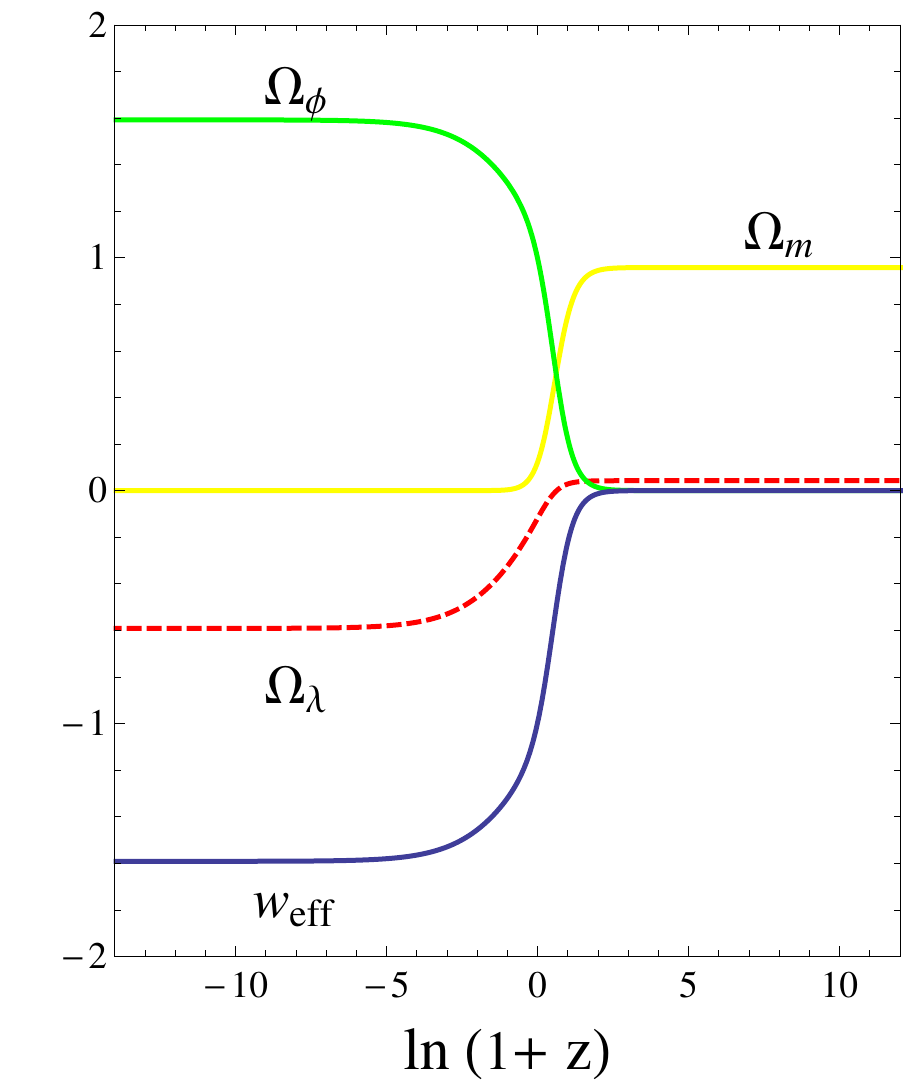}\label{
fig:beta_pot_parameters_1}}
\qquad
\subfigure[]{%
\includegraphics[width=7cm,height=5cm]{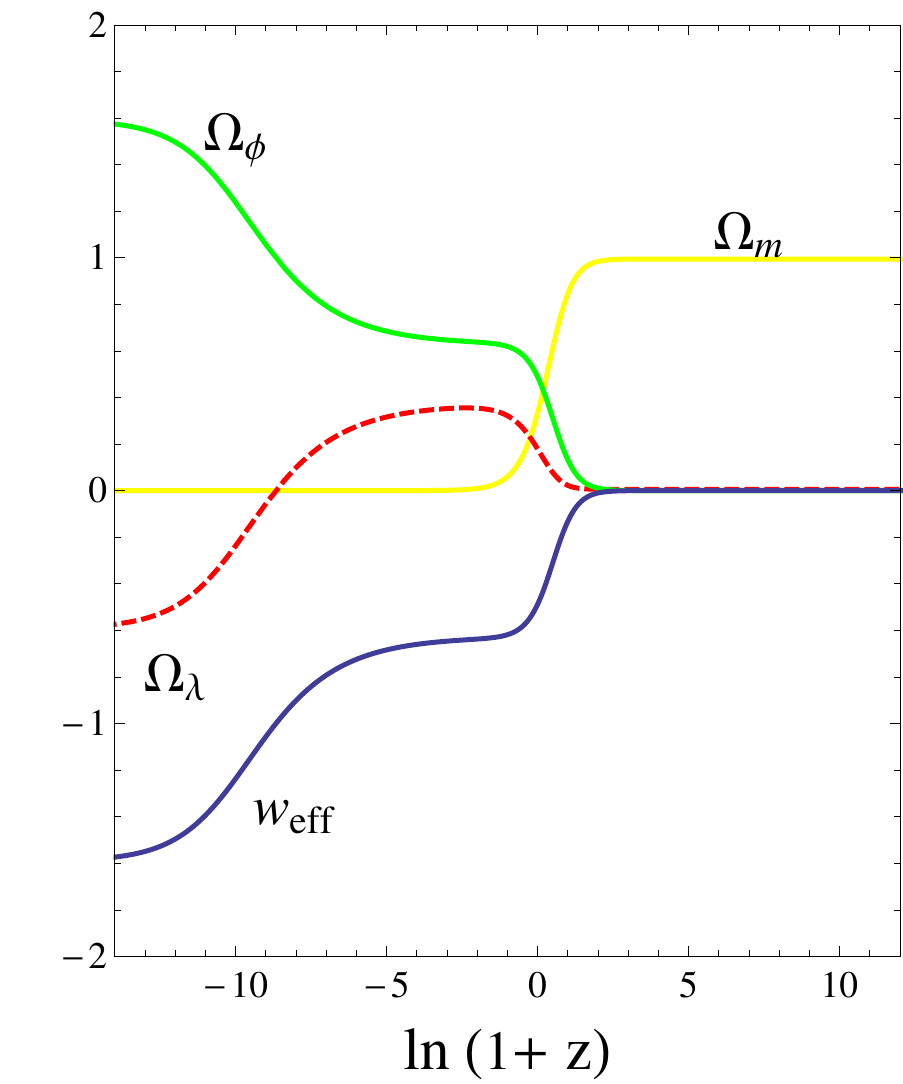}\label{
fig:beta_pot_parameters_2}}
\caption{
{\it{The evolution of the density parameters
$\Omega_m$, $\Omega_{\phi}$, $\Omega_\lambda$, as well as of the effective
equation-of-state parameter $w_{\rm eff}$, as functions of the redshift, for the
case of ppotential  $  V(\phi) = \alpha (1 + \beta \phi^2 )^{-2}$, with $\xi= - \frac{16
\beta}{\alpha \kappa^2}=0.3$ and for $w=0$. The left and right graphs correspond to
different initial conditions.
}}}
\label{fig:beta_pot_parameters}
\end{figure}

In order to see these features in an alternative way, in
Fig.~\ref{fig:beta_pot_parameters} we depict the evolution of the various density
parameters, as well as of the effective
equation-of-state parameter $w_{\rm eff}$, as functions of the redshift, for two choices
of the initial conditions. Qualitatively there are thus two possible cosmic evolutions.
The first one is depicted in the left panel of
Fig.~\ref{fig:beta_pot_parameters}, where a transition from matter domination to phantom
behavior is achieved. This is provided by trajectories directly connecting the critical
plane $A_2$ with point $A^-_{5+}$. The second evolution class is depicted in the right
panel of Fig.~\ref{fig:beta_pot_parameters}, where after matter domination a long lasting
finite period of quintessence-like acceleration is attained before a transition to the
final phantom domination happens. This scenario is very interesting because the
unstable accelerating scaling solution can be used to alleviate the cosmic coincidence
problem.
The future destiny of the universe is however a phantom dominated solution, which could
lead to a Big Rip singularity in a finite time. Note that a current phantom-like dark energy scenario appears more plausible than a quintessence-like in light of the reported preference for the normal neutrino mass hierarchy from cosmological data~\cite{Vagnozzi:2017ovm,Vagnozzi:2018jhn}.
Note also that in Fig.~\ref{fig:beta_pot_parameters} the relative energy densities are
again not constrained in the interval $[-1,1]$. As mentioned above this is an issue
related to cosmological models where an effective interaction between the
components sourcing the field equations is present \cite{Quartin:2008px,Tamanini:2015iia}.


\section{Cosmological implications}
\label{sec:cosmological_implications}

In the previous section we performed a detailed phase-space and stability  analysis for
the scenario of mimetic gravity with the inclusion of a scalar-field potential. In this
section we discuss the physical implications of the analysis.
The cosmological dynamics of the mimetic gravity exhibits a very rich phenomenology,
including for example matter to dark energy transitions, phantom behavior and
accelerated scaling solutions. Apart from being in agreement with observations, these
solutions can be in fact used to alleviate some of the problems afflicting modern
cosmology, such as the cosmic coincidence problem.

In the case of a general potential, and irrespectively of its form, the cosmological
behavior of mimetic gravity will always admit both matter dominated solutions, and
dark-energy dominated solutions mimicking a cosmological constant, or quintessence-like,
or phantom-like behavior. Furthermore, the
phase space presents scaling solutions too, in which the matter
and dark-energy density parameters are of the same order and thus they can offer an
alleviation to the coincidence problem. Hence, one can easily describe the universe
history in agreement with observations, namely the successive sequence of radiation,
matter, and dark-energy eras.

Specifying the potential to the case of an inverse square form, not only yields
accelerating scaling solutions, which can be used to solve the cosmic coincidence
problem,
but it also provides an early time matter-dominated state, which in the case of $w=1/3$
can be used to well characterize the radiation era, while for $w=0$ it corresponds to the
dust matter epoch. In fact, choosing proper initial conditions of the universe, the
cosmic
dynamics of mimetic gravity with this potential, and with suitable parameter choices,
will
lead to a universe which  starts from a radiation phase, it enters into a
matter-dominated epoch, and then it transits to a scaling solution with $w_{\rm eff}
\simeq -0.7$, in agreement with observations. This scenario is thus not only
mathematically simple to be analyzed (it yields a two-dimensional dynamical system), but
it is also phenomenologically very powerful.

The cosmological scenario is different for mimetic gravity with an exponential
or power-law potential. In this case there is no accelerating scaling solution and there
is no finite final attractor in the phase space.
The observed transition from matter to dark energy domination can still however be
correctly described since a finite period of cosmological constant behavior can still be
attained after a long-lasting matter solution.
An expansion history similar to $\Lambda$CDM cosmology can thus still be achieved,
although in the far future the universe may be lead to a sudden Big Rip singularity.

In the case of the potential $  V(\phi) = \alpha (1 + \beta \phi^2 )^{-2}$, we found two
possible qualitative cosmic evolutions. In the first one we obtained a matter-dominated
early time solution, followed by a phantom-like late-time accelerated phase. In the
second class of cosmic evolutions we found that between the early-time matter
domination and the future phantom behavior, there is a long lasting finite period
described by an accelerating scaling solution, which can actually be used to alleviate
the
cosmic coincidence problem. Choosing suitable model parameters one can also obtain
$w_{\rm
eff} \simeq -0.7$ during the scaling regime, in agreement with
observations~\citep{ade:2015xua}.
This model can thus both alleviate  the cosmic coincidence problem and provide a
well-behaved early-time dynamics, similarly to the inverse square potential.
It nevertheless leads to a different cosmological scenario where the present state of the
universe is described by an accelerating scaling solution, but in the future a transition
to a phantom regime will happen and the universe may approach a Big Rip
singularity.

\section{Conclusions}
\label{sec:conclusion}

Mimetic gravity has emerged as an interesting alternative to general relativity. 
Within this theory, the conformal degree of freedom of gravity is isolated in a 
covariant way through a singular disformal transformation: the resulting dynamics changes 
and an   effective dark matter component   appears    on cosmological scales.
Various works have
previously investigated background cosmological solutions in mimetic gravity, finding
that appropriate choices of the potential for the mimetic field lead to appealing
cosmological solutions, which can reproduce expansion histories in agreement
with observational data without the need for additional dark matter or dark energy fluids.

In this work we have performed a dynamical-systems analysis of mimetic gravity, which
has allowed us to study the cosmological dynamics of the theory. We have focused on both
general potentials for the mimetic field, as well as on a set of well-motivated specific
choices. From the point of view of the
expansionary history, our analysis suggests that the potential whose solutions are most
appealing is the inverse square potential. The corresponding solutions possess an
early-time radiation phase, followed successively by matter era and late-time
accelerating scaling solution, which can alleviate the coincidence problem.
We mention that this alleviation of the coincidence problem is obtained without imposing
an interaction between  dark-matter and dark-energy sectors by hand, as it is the usual
approach \citep{amendola:1999er}, but it arises from the scenario of
mimetic gravity itself. Therefore, with this choice of potential, mimetic gravity is in
agreement with observations, and it provides the correct phenomenology at
early times, as opposed to simpler scalar-field models of dark energy where a
stiff-matter
dominated early-time solution is usually attained.
Moreover, it should be remarked that the inverse square potential is also very well
motivated from a high-energy ultraviolet completion point of view, aside from being
renormalizable, which lends even more to the attractiveness of the model.

In summary, mimetic gravity with an inverse square potential yields:

\begin{itemize}
\item A unified description of both dark matter and dark energy with a single
scalar field;

\item A well behaved early-time phenomenology;

\item The successive sequence of radiation, matter, and dark-energy eras, with
transitions in agreement with observations;

\item Late time accelerated scaling solutions that can alleviate the cosmic
coincidence problem;

\item A scalar-field potential with interesting theoretical features.

\end{itemize}

In conclusion in our work we have shown that the cosmological dynamics of mimetic gravity
renders the theory a viable and interesting candidate to explain the universe's
history. At the same time, it is important and timely to go beyond the background
analysis
and analyze and understand structure formation within this scenario, as well as provide a
full Markov Chain Monte Carlo analysis, comparing the model with observational data. It
will also be important to confirm or discard the raised issues concerning instabilities,
in order to ascertain whether or not the theory of mimetic gravity is really viable. We
leave these issues to future projects.

\acknowledgments
 J.D. acknowledges the support of Associate program of IUCAA.
N.T.~acknowledges partial support from the Labex P2IO and an Enhanced Eurotalents
Fellowship. S.V. ~acknowledges support by the Vetenskapsr\aa det (Swedish Research
Council)
through contract No. 638-2013-8993 and the Oskar Klein Centre for Cosmoparticle Physics.
N.T. and S.V. wish to thank the NORDITA cosmology program ``Advances in theoretical
cosmology in light of data'' and in particular Martina Gerbino, scientific organizer for
the first week of the program, where discussions leading to this work were conducted.
  E.N.S. desires to thank Institut de Physique Th\'{e}orique,
Universit\'e Paris-Saclay and Laboratoire de Physique Th\'eorique, Univ. Paris-Sud,
Universit\'e Paris-Saclay, for the hospitality during the last stages of this project.
This article is based upon work from COST Action ``Cosmology and Astrophysics Network
for Theoretical Advances and Training Actions'', supported by COST (European Cooperation
in Science and Technology).

\begin{appendix}

\section{Appendix: stability analysis for $A_1$, $A_2$}
 \label{A1A2stability}

In this Appendix we analyze the stability of the curves of critical points   $A_1$ and
$A_2$ presented in Table \ref{tab:c_pts_general}. As it is known a  set of  non-isolated
critical points with $N$ vanishing eigenvalues is called ``normally hyperbolic set of
dimension $N$''. The $N$-dimensional eigenspace spanned by the eigenvectors
corresponding to the vanishing eigenvalues (eigenvalues with
vanishing real part) determine the direction of the critical
set. The stability of this set is therefore determined by the behavior of trajectories on
the eigenspace spanned by eigenvectors corresponding to the non-vanishing eigenvalues
(eigenvalues with non-vanishing real part).
Thus, the stability depends on the signature of nonvanishing eigenvalues.

In order to properly determine the stability behavior of sets $A_1$ and $A_2$, we shall
examine the  $w \neq 0$ and   $w=0$ separately.

\subsection{Case: $w\neq 0$}

In this case curves $A_1$ and $A_2$ are normally hyperbolic sets of critical points.
Hence, looking from the signature of non-vanishing eigenvalues, we can conclude
that $A_1$ behaves as a saddle and $A_2$ behaves as an unstable node.

\subsection{Case: $w=0$}

In this case the stability depends on the value of $\Gamma(z)$:

\begin{itemize}
\item  For potentials with $\Gamma(z)=\frac{3}{2}$, the $x$-axis  becomes a critical
line. Thus, curves $A_1$, $A_2$ are replaced by a single critical line, namely the
$x$-axis. Since the stability matrix of the $x$-axis has  eigenvalues  ($\frac{3}{2}$, 0)
it is normally hyperbolic. Hence, the $x$-axis behaves as an unstable node.

\item  For potential with $\Gamma(z) \neq \frac{3}{2}$, the plane $y=0$
becomes a critical plane. Therefore, curves $A_1$, $A_2$ are replaced by this plane.
The plane $y=0$ is normally hyperbolic of dimension 2, with eigenvalues ($\frac{3}{2},
0,0$). Thus, the stability depends on the signature of non-vanishing eigenvalues. Hence,
the plane $y=0$ is behaving as an unstable node.

\end{itemize}

We close this Appendix by mentioning that we have indeed verified that in the above cases
the center manifold is actually the critical set itself. Hence, the behavior of
trajectories on the center manifold cannot provide the stability
of the critical set. Nevertheless, the stability is completely determined by
the behavior of trajectories on the eigenspace, spanned by eigenvectors corresponding to
the remaining non-vanishing eigenvalues. Thus, the stability depends on the signature of
the non-vanishing eigenvalues.

The stability results of curves $A_1$ and $A_2$  are summarized in Table
\ref{tab:eigen_general}.

\section{Appendix: stability analysis for  $A_{3\pm}$}
\label{CMTA3}

In this Appendix, we apply the center manifold method  \cite{Leon2011} in order to
study the stability of the non-hyperbolic points $A_{3\pm}$ presented in Table
\ref{tab:c_pts_general}. We first translate the point  $A_{3+}\,(0,1,0)$ to the origin
via the transformation $x\rightarrow x$,
$y\rightarrow y+1$,
 $z\rightarrow z$.
We then introduce a new set of variables $(X,Y,Z)$, defined in terms of the original set
of variables $(x,y,z)$ as
\[\left(\begin{array}{c}
X\\
Y\\
Z \end{array} \right)=\left(\begin{array}{ccc}
-2   & 0 &  \frac{1}{\sqrt{3}}  \\
1    &1   & -\frac{\sqrt{3}}{6}\\
0    & 0 &  1\\ \end{array} \right) \left(\begin{array}{c}
x\\
y\\
z \end{array} \right) \,. \]
In terms of these new set of variables, the system of equations can   be re-written as
\[\left(\begin{array}{c}
X'\\
Y'\\
Z' \end{array} \right)=\left(\begin{array}{ccc}
-3  & 0 & 0  \\
0  & -3 (w+1) & 0\\
0  & 0 & 0   \end{array} \right) \left(\begin{array}{c}
X\\
Y\\
Z \end{array} \right)+\left(\begin{array}{c}
g_1\\
g_2\\
f \end{array} \right),\]
where $f,\,g_1,\,g_2$ are polynomials of degree greater than 2 in $(X,\,Y,\,Z)$, with
\begin{eqnarray}
f(X,Y,Z)&=&-\frac{1}{4}\, \left( 2\, \Gamma(Z)-3 \right)  \left( 2\,
\sqrt {3}\,X+2\,\sqrt {3}\,Y+2\,\sqrt {3}-Z \right) {Z}^{2}
\end{eqnarray}
and with $g_1$, $g_2$ not explicitly presented here due to their
lengths.
The center manifold is locally represented by
\begin{equation}
\left\lbrace (X,Y) : X=h_1(Z), Y=h_2(Z), h_i(0)=0,  Dh_i(0)=0 , i=1,2 \right\rbrace,
\end{equation}
where $h_1$, $h_2$ are approximated as
\begin{align}
h_1(Z)=a_2 Z^2+a_3 Z^3+\mathcal{O}(Z^4),\\
h_2(Z)=b_2 Z^2+b_3 Z^3+\mathcal{O}(Z^4),
\end{align}
respectively. Due to the invariant property of the center manifold the vector function
\[\mathbf{h}=\left(\begin{array}{c}
h_1\\
h_2\ \end{array} \right)
\]
has to satisfy a quasilinear partial differential equation given by
\begin{equation}\label{quasi_C}
D \mathbf{h(S)}\left[A S+\mathbf{F}(S,\mathbf{h}(S))\right]-B
\mathbf{h}(S)-\mathbf{g}(S,\mathbf{h}(
S))=\mathbf{0} \,,
\end{equation}
with \[\mathbf{g}=\left(\begin{array}{c}
g_1\\
g_2 \end{array} \right),~~~~~ \mathbf{F}=f, ~~~~~B= \left(\begin{array}{cc}
 -3 & 0 \\
 0 &-3(w+1) \end{array} \right),~~~~~ A=0. \]
On substituting $A$, $\textbf{h}$, $\mathbf{F}$, $B$, $\mathbf{g}$ into equation
\eqref{quasi_C}
and equate the coefficients of all powers of $Z$ to zero, we obtain the constants $a_2$,
$a_3$, $b_
2$, $b_3$ as
\begin{gather}
    a_2=-\frac{\Gamma (0)}{6}+\frac{1}{3} \,,\quad
    a_3=-\frac{49\sqrt{3}}{144}+\frac{\sqrt{3}}{18}
\Gamma(0)\left[7-2\Gamma(0)\right]-\frac{1}{6}\,
\Gamma_z(0)\nonumber  \\
    b_2=-\frac{1}{24}\,,\qquad~~~   b_3=-\frac{\sqrt{3}}{144}\,\left[4\,\Gamma(0)-7
\right].
\end{gather}
The dynamics of the reduced system is ultimately determined by the equation
\begin{equation}
Z'=A\,Z+\mathbf{F}(Z,\mathbf{h}(Z)),
\end{equation}
and hence
\begin{align}
S'=\sqrt{3}\left[-\Gamma(0)+\frac{3}{2}\right]
Z^2+\left\{\frac{1}{2}\left[\Gamma(0)-\frac{3}{2}\right]-\sqrt{3} \Gamma_z(0)\right\}\,Z^3
+\mathcal{O}(Z^4).
\end{align}
Therefore, for $\Gamma(0) \neq \frac{3}{2}$, point $C_{3+}$ is saddle. However, if
$\Gamma(0)=\frac{3}{2} $ then the next terms in the expansion must be considered, in
which case the point is stable if   $ \Gamma_z(0)>0$.

A similar analysis for point $A_{3-} (0,-1,0)$ shows that it is saddle
for  $\Gamma(0) \neq \frac{3}{2}$, however for  $\Gamma(0)=\frac{3}{2}$ this point is
stable if $  \Gamma_z(0)<0$.

The stability results of points $A_{3\pm}$ are summarized in Table
\ref{tab:eigen_general}.

\end{appendix}

\bibliographystyle{JHEP}

\providecommand{\href}[2]{#2}\begingroup\raggedright\endgroup

\end{document}